\documentclass[floats,floatfix,twocolumn,superscriptaddress,nofootinbib,showpacs]{revtex4-1}

\pdfoutput=1
\usepackage{amsfonts,amssymb,amsmath}
\usepackage[usenames,dvipsnames]{xcolor}
\usepackage[unicode,colorlinks=true,linkcolor=blue,urlcolor=blue,citecolor=gray,pdfusetitle]{hyperref}
\usepackage[all]{hypcap}
\usepackage{graphicx}
\usepackage{xspace}
\usepackage{tikz}
\usetikzlibrary{shapes,arrows}
\usetikzlibrary{matrix}
\usetikzlibrary{shapes.multipart}
\usetikzlibrary{er,positioning}
\usepackage{orcidlink}
\usepackage[normalem]{ulem} %for \sout
\usepackage{url}
\usepackage{color,soul}
\usepackage[caption=false]{subfig}
\usepackage[T1]{fontenc}
\usepackage[utf8]{inputenc}
\usepackage{microtype}
\newcommand{\Flatiron}{\affiliation{Center for Computational Astrophysics, Flatiron Institute, 162 5th Ave, New York, NY 10010, United States}}
\newcommand{\StonyBrook}{\affiliation{Department of Physics and Astronomy, Stony Brook University, Stony Brook NY 11794, United States}}
\newcommand{\UMS}{\affiliation{Department of Physics and Astronomy,
  The University of Mississippi, University, MS 38677, United States}}
\newcommand{\MIT}{\affiliation{LIGO Laboratory and Kavli Institute for Astrophysics and Space Research, Massachusetts Institute of Technology, Cambridge, Massachusetts 02139, USA}}

\newcommand{\pd}{\partial}

\newcommand{\nn}{\nonumber}

\newcommand{\pont}{{}^{*}\!RR}
\newcommand{\dual}{\,{}^*\!}
\newcommand{\He}{\mathcal{H}}
\newcommand{\RL}{{\mathrm{R}, \mathrm{L}}}
\newcommand{\Ro}{\mathrm{R_{obs}}}
\newcommand{\io}{\iota_\mathrm{obs}}
\newcommand{\RR}{\mathrm{R}}
\newcommand{\Lo}{\mathrm{L_{obs}}}
\newcommand{\LL}{\mathrm{L}}

\begin{document}

\title{Constraining gravitational wave amplitude birefringence and Chern-Simons gravity with GWTC-2}

\author{Maria Okounkova\,\orcidlink{0000-0001-7869-5496}}
\email{mokounkova@flatironinstitute.org}
\Flatiron

\author{Will M. Farr\,\orcidlink{0000-0003-1540-8562}}
\email{wfarr@flatironinstitute.org}
\StonyBrook{} \Flatiron{}

\author{Maximiliano Isi\,\orcidlink{0000-0001-8830-8672}}
\email{maxisi@mit.edu}
\thanks{NHFP Einstein fellow}
\Flatiron
\MIT

\author{Leo C. Stein\,\orcidlink{0000-0001-7559-9597}}
\email{lcstein@olemiss.edu}
\UMS

% Because hyperref only gets the *last* author, we need to be explicit.
\hypersetup{pdfauthor={Okounkova et al.}}

\date{\today}

\begin{abstract}
We perform a new test of general relativity (GR) with signals from GWTC-2, the LIGO and Virgo catalog of gravitational wave detections. We search for the presence of amplitude birefringence, in which left versus right circularly polarized modes of gravitational waves are exponentially enhanced and suppressed during propagation. Such an effect is present in various beyond-GR theories but is absent in GR. We constrain the amount of amplitude birefringence consistent with the data through an opacity parameter $\kappa$, which we bound to be $\kappa \lesssim 0.74 \textrm{ Gpc}^{-1}$. Our constraint is derived under an assumption that all GWTC-2 events have a common distance. This result for $\kappa$ is statistically significant, with a Jensen-Shannon divergence of $7 \times 10^{-2}$ bits compared to an uninformative distribution on $\kappa$. We then use these theory-agnostic results to constrain Chern-Simons gravity, a beyond-GR theory with motivations in  quantum gravity. We bound the canonical Chern-Simons lengthscale to be $\ell_0 \lesssim 1.0 \times 10^3$ km, in agreement with other long-distance measurement results. 
\end{abstract}

\maketitle

\section{Introduction}

At some length scale, Einstein's theory of general relativity (GR) must break down and be reconciled with quantum mechanics in a beyond-GR theory of gravity. Gravitational waves (GWs) from binary black hole (BBH) mergers, such as those recently detected by LIGO \cite{TheLIGOScientific:2014jea} and Virgo~\cite{TheVirgo:2014hva} could contain signatures of beyond-GR effects,  which has motivated significant efforts to test GR with LIGO and Virgo detections~\cite{LIGOScientific:2018mvr,LIGOScientific:2019fpa, Isi:2019aib, Nair:2019iur, Isi:2019asy, Abbott:2020niy, Abbott:2020jks}.

One particular beyond-GR effect to study is \textit{amplitude birefringence}: in several beyond-GR theories, when GWs propagate from the source to the detector, the amplitudes of left versus right polarized modes are exponentially enhanced or suppressed, a parity-violating effect. This effect is absent in general relativity. The strength of this effect is governed by a universal \textit{opacity parameter}, $\kappa$, whose value is zero in GR.\footnote{Note that the symbol $\kappa$ is often used in the literature for the  constant $(16\pi G)^{-1}$ in the Einstein-Hilbert action of GR.  In this study we use it solely to denote the amplitude birefringence opacity parameter.}  In order to study a specific beyond-GR theory, the value of $\kappa$ can be mapped onto the parameters governing that theory. In particular, theories that exhibit amplitude birefringence include Chern-Simons gravity~\cite{Alexander:2009tp}, ghost-free scalar-tensor theories~\cite{Crisostomi:2017ugk}, symmetric teleparallel equivalents of GR~\cite{Conroy:2019ibo}, and Ho\v{r}ava-Lifshitz gravity~\cite{Horava:2009uw} (a thorough theoretical review of amplitude birefringence in beyond-GR theories is provided in~\cite{Zhao:2019xmm}).

Indeed, recent studies have looked for GW amplitude birefringence within the first LIGO and Virgo GW transient catalog GWTC-1~\cite{LIGOScientific:2018mvr}. Recently, both Wang et al.~\cite{Wang:2020cub} and Yamada et al.~\cite{Yamada:2020zvt} analyzed the presence of amplitude birefringence in GWTC-1 by comparing data against GW \textit{template waveforms} that included birefringence effects, finding no evidence of parity violation. 

%%\Note{Maybe write about the Shao~\cite{Shao:2020shv} and Kostaleky~\cite{Kostelecky:2016kfm} and Mewes~\cite{Mewes:2019dhj} (the corresponding theory paper) papers here as the referee wants, but Shao studies the time delay and velocity birefringence and bounds on some general gravitational Lorentz invariance violation parameters}.}  

In this study, we perform a novel, simpler test of GR by constraining GW amplitude birefringence using the confident BBH detections in GWTC-2, the second LIGO-Virgo catalog~\cite{Abbott:2020niy, Abbott:2019ebz, GWOSC:PE}\footnote{Note that GWTC-2 contains GWTC-1~\cite{LIGOScientific:2018mvr}, the first LIGO and Virgo catalog, as a subset.}. As described in Alexander et al.~\cite{Alexander:2007kv}, amplitude birefringence affects the distribution of \textit{observed} BBH inclination angles, either favoring all face-on or face-off detections (depending on the sign of $\kappa$). We thus use the reported posteriors for BBH inclination angles in GWTC-2 to constrain GW amplitude birefringence, placing bounds on $\kappa$, the univeral opacity parameter. This is a simpler, faster analysis, as it does not require using template waveforms and performing additional parameter estimation.

As a specific application, we use our limit on amplitude birefringence to constrain non-dynamical Chern-Simons gravity, a parity-violating beyond-GR effective field theory with origins in string theory, loop quantum gravity, and inflation~\cite{Alexander:2009tp, Green:1984sg, Taveras:2008yf, Mercuri:2009zt, Weinberg:2008hq}. Indeed, previous works have addressed the possibility of detecting Chern-Simons amplitude birefringence with GW detectors~\cite{Nojiri:2019nar, Zhao:2019szi, Alexander:2007kv, Yunes:2010yf, Yunes:2008bu,Yagi:2017zhb}, and in this study we perform such a measurement on real GW data.

In Sec.~\ref{sec:theory}, we give an overview of the observational effects of amplitude birefringence on GW detections, and outline our methods for measuring this effect. We then use GWTC-2 to bound the amount of amplitude birefrigence in BBH signals in Sec.~\ref{sec:observed_constraint}. In Sec.~\ref{sec:cs_implications}, we consider these results in the context of Chern-Simons gravity, and bound the canonical Chern-Simons lengthscale. We conclude in Sec.~\ref{sec:conclusion}.  We set $G = c = 1$ throughout. $H_0$ refers to the present day value of the Hubble parameter, with dimensions of $[H_0] = L^{-1}$, and $z$ refers to cosmological redshift.

\section{Theoretical background and methods}
\label{sec:theory}

\subsection{Gravitational wave amplitude birefringence}
\label{sec:observational_effects}

In GR, for the dominant $(2,\pm 2)$ angular mode of non-precessing compact binary inspirals, the ratio of the gravitational wave strain $h$, in right $h_\RR$, versus left $h_\LL$, circularly polarized modes is purely a function of the inclination angle of the binary, of the form
\begin{align}
\label{eq:GRPolarization}
    \left(\frac{h_\RR}{h_\LL}\right)_\mathrm{GR} = \left(\frac{1 + \cos \iota}{1 - \cos \iota}\right)^2\, .
\end{align}
Here, the inclination $\iota$ is the angle from the total angular momentum of the binary to the line of sight of the observer.  In terms of the plus, $h_+$, and cross, $h_\times$, polarizations, the circular polarizations are given by $h_{\RR, \LL} = h_+ \pm i h_\times$. A system with $\cos \iota = 1$ has power purely in $h_\RR$, and is \textit{face-on}, while one with $\cos \iota = -1$ has power purely in $h_\LL$ and is \textit{face-off}.  Thus
\begin{align}
    \textrm{pure } h_\RR &\Longleftrightarrow \cos \iota = +1 \Longleftrightarrow \textrm{face-on}\,, \\
    \textrm{pure } h_\LL &\Longleftrightarrow \cos \iota = -1 \Longleftrightarrow \textrm{face-off}\,.
\end{align}
We assume that the universe is homogeneous and isotropic at cosmological scales, and that gravitational physics does not have any preferred direction.  This implies that the underlying distribution for $\cos \iota$ is flat, meaning no preference for face-on versus face-off events.

The picture in Eq.~\eqref{eq:GRPolarization} changes in beyond-GR theories that exhibit amplitude birefringence. In this case, the amplitudes of left- versus right-polarized modes are exponentially enhanced and suppressed during propagation, leading to an expression of the form
\begin{align}
\label{eq:hKappa}
    \left(\frac{h_\Ro}{h_\Lo}\right)_\mathrm{Biref} = \frac{e^{-d_C \kappa}(1 + \cos \iota)^2}{e^{d_C \kappa}(1 - \cos \iota)^2}\,.
\end{align}
Here, $d_C$ is the comoving distance to the source (with units of length $L^1$), and $\kappa$ is an \textit{opacity parameter} with units of $L^{-1}$ that governs the strength of the birefringence. Note that $\kappa = 0$ is consistent with GR. The above expression uses the comoving distance, as the birefringence effect accumulates with the distance traveled as `experienced' by the graviton.

Throughout this study, we will assume that $\kappa d_C \ll 1$, that is, beyond-GR effects are small enough that the effective field theory is valid. Note that while Eq.~\eqref{eq:hKappa} assumes that $\kappa$ is a constant, in some beyond-GR scenarios (including potentially more complicated Chern-Simons scalar field profiles than those considered in Sec.~\ref{sec:cs_implications}), the strength of the birefringence may have a more complicated dependence on the comoving distance $\kappa = \kappa(d_C)$. However, if we expand this dependence to linear order, $\kappa(d_C) = \kappa_0 + \mathcal{O}(d_C)$, then we can treat Eq.~\eqref{eq:hKappa} as correct to linear order for every theory.

When performing tests of general relativity with GW amplitude birefringence, one of our goals is to map values or constraints on $\kappa$, a `universal' quantity, to \textit{specific} beyond-GR theories. In some instances of certain beyond-GR theories of gravity, Eq.~\eqref{eq:hKappa} with a constant value of $\kappa$ is the precise form of the amplitude birefringence. In other instances, however, $\kappa$ may be dependent on $d_C$, making the problem more difficult. However, we can treat Eq.~\eqref{eq:hKappa} with a constant value of $\kappa = \kappa_0$ as the \textit{leading} term in an expansion in $d_C$, of the form $\kappa\left( d_C \right) = \kappa_0 + \mathcal{O}\left( d_C \right)$. For the remainder of the paper, we assume a constant value of $\kappa$.

In traditional GW parameter estimation, however, we do not have access to the \textit{true} value, $\cos \iota$, of the inclination angle, but rather observe some \textit{effective} value, $\cos \io$. Thus, from Eq.~\eqref{eq:hKappa}, in the presence of amplitude birefringence, we would measure a ratio
\begin{align}
\label{eq:Inclinations}
    \frac{1 + \cos \io}{1 - \cos \io} = \frac{e^{-d_C \kappa/2}(1 + \cos \iota)}{e^{d_C \kappa/2}(1 - \cos \iota)}\,.
\end{align}

Let us think about how amplitude birefringence would affect the values $\cos \io$ for multiple events. Statistical isotropy of BBH orientation requires that $p(\cos \iota)$, the distribution on the true inclination angle over the population of BBH mergers, be flat. The \emph{observed} distribution of inclinations is influenced by selection effects, but to a very good approximation these are independent of the \emph{sign} of $\cos \iota$ \citep{Abbott:2020niy,Abbott:2020gyp}.  Thus, if there are no beyond-GR effects and $\kappa = 0$, we expect to see an equal number of face-on and face-off events. Meanwhile, if $\kappa > 0$, we will preferentially measure $\cos \io \sim -1$ for isotropically distributed events. In other words, we will preferentially see more face-off, rather than face-on mergers. Similarly, if $\kappa < 0$, we will preferentially see more face-on mergers. Thus, we expect $p(\cos \io)$ to not be symmetric about zero.

\subsection{Measuring amplitude birefringence}
\label{sec:BirefringenceMeasurement}

We can now use the fact that GW amplitude birefringence changes the  distribution of observed inclination angles (to prefer either more face-on or face-off events) to quantify and constrain amplitude birefringence with gravitational wave data.

A simple method is to use an \textit{asymmetry statistic} to quantify the number of observed face-on versus face-off events. Let us define  the on/off (or right/left) asymmetry statistic
$\Delta$, in the range $-1 \le \Delta \le +1$, as
\begin{align}
  \label{eq:DeltaDef}
  \Delta \equiv \frac{N(\cos\io > 0) - N(\cos\io < 0)}{N}
  \,,
\end{align}
where $N$ is the total number of GW observations, $N(\cos\io > 0)$ is the number of face-on observations, and $N(\cos\io < 0)$ is the number of face-off observations. An underlying distribution on $\cos\iota$ and the birefringence
effect in Eq.~\eqref{eq:Inclinations} will thus induce a distribution on $\Delta$. Working solely with a quantity such as $\Delta$ provides a robust framework for many beyond-GR theories, and does not require making assumptions about the underlying theory, as is done when producing template waveforms.

In practice, when analyzing gravitational wave data, we have, for each GW event, a posterior \textit{distribution} on $\cos \io$, $p(\cos \io \mid d)$ (where $d$ corresponds to the data), rather than a single value. Thus, our goal is to go from individual $p(\cos \io \mid d)$ distributions to an overall distribution $p(\Delta \mid d)$.

For each GW event we consider, let us take $n_\mathrm{samp} = 1024$ samples from the posterior distribution $p(\cos \io \mid d)$. Then, for each event, let us compute a scalar
\begin{align}
\label{eq:deltaDef}
\delta \equiv n_\mathrm{samp}(\cos \io < 0)/n_\mathrm{samp},
\end{align}
quantifying the number of negative (face-off) samples, and $(1 - \delta)$ quantifying the number of face-on samples. Then, given a value of $-1 \le \Delta \le +1$, we can compute a likelihood of the form
\begin{align}
\label{eq:DeltaProduct}
    p(d \mid \Delta) = \prod_{i = 1}^{N} \delta_i \left( \frac{1 - \Delta}{2}\right) + (1 - \delta_i) \left(\frac{1 + \Delta}{2} \right)
\end{align}
where the product is over the GW events, and $\delta_i$ corresponds to $\delta$ (Eq.~\eqref{eq:deltaDef}) for each event.

Note that we can express $\Delta$ in terms of $p(\cos \io \mid d_C, \kappa)$ as
\begin{align}
\label{eq:DeltaIntegralIndentity}
    \int_{-1}^0 \mathrm{d} \cos \io \, p\left( \cos \io \mid d_C, \kappa \right) = \frac{1-\Delta}{2}\,.
\end{align}

Our goal now is to map the resulting $\Delta$ onto a physical parameter, $\kappa$. We can achieve this by substituting an expression for $p(\cos \io \mid d_C, \kappa)$ into Eq.~\eqref{eq:DeltaIntegralIndentity} and evaluating the integral. Using the chain rule, we can express
\begin{align}
    p(\cos \io \mid d_C, \kappa) = \frac{d \cos \iota}{d \cos \io} p(\cos \iota \mid d_C, \kappa)\,.
\end{align}
We expect the true binary black hole inclination angle to be isotropically distributed and independent of $\kappa$ and $d_C$, of the form of a flat distribution $\cos \iota \sim \mathcal{U}(-1, 1)$.
  Thus, we obtain
\begin{align}
    p(\cos \io \mid d_C, \kappa) = \frac{1}{2}\frac{d \cos \iota}{d \cos \io}\,.
\end{align}
Using Eq.~\eqref{eq:Inclinations} to evaluate $d \cos \iota / d \cos \io$, we obtain
\begin{align}
    p(\cos \io \mid d_C, \kappa) = \frac{1}{2} \left( \cosh \frac{d_C \kappa}{2} - x_\mathrm{obs} \sinh \frac{d_C \kappa}{2} \right)^{-2}\,.
\end{align}
Finally, plugging this expression into Eq.~\eqref{eq:DeltaIntegralIndentity} and evaluating the integral, we obtain an expression for $\kappa$ in terms of $\Delta$,
\begin{align}
\label{eq:KappaDelta}
    \kappa = \frac{1}{d_C} \log\left[ \frac{1 + \Delta}{1-\Delta} \right]\,.
\end{align}
Thus, given a value of $\Delta$, we can now map onto a physical value of $\kappa$.

It is appropriate to match the value of $\Delta$ inferred from the data to the effect of $\kappa$ on the astrophysical population rather than the \emph{selected} population (events that pass some detection threshold) for the following reason.  Selection effects are, to a very good approximation, independent of the \emph{sign} of $\cos \io$ \citep{LIGOScientific:2018mvr,Abbott:2020niy}; due to this symmetry, the same fraction of the population of mergers will be detectable for \emph{any} value of $\Delta$ in our simplified model where the distribution of $\cos \io$ is piecewise-flat.  The usual factor correcting for selection effects, conventionally written $\alpha \left(\Delta \right)$  \citep{Mandel:2018mve}, appearing in the denominator of the likelihood is therefore constant.  Our analysis, ignoring the constant $\alpha$ factor, infers the true \emph{population} value of $\Delta$; and it is therefore appropriate to match inferred $\Delta$ values to the actual effect on the population from $\kappa$ rather than the \emph{selected} population.

We can simplify the above analysis if we assume that all mergers come from the same distance. The effect of birefringence on the observed inclination depends on the product of the opacity parameter and the comoving distance to each event, and a full analysis would take account of the varying distances to the events in GWTC-2, which we perform in Appendix~\ref{sec:appendix_distance}.  To obtain an approximate constraint averaging over BBH detections using a simple counting argument, however, we approximate a common comoving distance, $d_C$, for all events.

Assuming that $\kappa d_C$ is the same for all observations and that $\cos \iota \sim \mathcal{U}(-1, 1)$ we get the expected value
\begin{align}
  \label{eq:DeltaKappa} \hat{\kappa} =
\frac{1}{d_C} \log\left[ \frac{1 + \hat{\Delta}}{1-\hat{\Delta}} \right]\,.
\end{align}
An error in our distance assumption will come in as an effect that is of order $d_C^{-2}$. If there's a small error $\epsilon$ in the assumed distance, then we will obtain
\begin{align}
\hat{\kappa} &= \frac{1}{d_C + \epsilon} \log\left[ \frac{1 + \hat{\Delta}}{1-\hat{\Delta}} \right] \\
\nn &= \frac{1}{d_C} \log\left[ \frac{1 + \hat{\Delta}}{1-\hat{\Delta}} \right] - \frac{\epsilon}{d_C^2} \log\left[ \frac{1 + \hat{\Delta}}{1-\hat{\Delta}} \right]\ldots \,.
\end{align}

\subsection{Additional considerations}
\label{sec:additional_considerations}

In this study we are considering beyond-GR modifications to a gravitational waveform as it propagates from the source to the detector. We do not, however, consider beyond-GR effects \textit{at the source itself}, which would change the phase of the waveform. This is in the spirit of the tests of general relativity presently performed by the LIGO and Virgo collaborations~\cite{LIGOScientific:2019fpa, Abbott:2020jks}, which consider beyond-GR modifications to the generation and the propagation of GWs independently. Future tests of GR should in deed consider \textit{both} source frame and propagation effects in beyond-GR theories. For certain theories which exhibit phase modification as well, such as Chern-Simons gravity, the comped source frame dynamics for systems such as binary pulsars do lead to a stronger constraint on the lengthscale governing the theory (cf. Sec.~\ref{sec:cs_implications}).

Birefringence also changes the signal \emph{amplitude} measured at the detector,
and therefore the inferred luminosity distance to the source, via
\begin{multline}
\frac{d_{L,\mathrm{obs}}}{d_L} = \\ \frac{\sqrt{1 + \cos^2 \iota}}{\sqrt{\left( 1 + \cos^2 \io \right) \cosh 2 \kappa d_C + 2 \cos \io \sinh 2 \kappa d_C}} \\
= 1 + \frac{\cos \io \left( \cos^2 \io - 5 \right)}{2 \left( 1 + \cos^2 \io \right)} \kappa d_C + \mathcal{O}\left( \kappa d_C \right)^2 \,,
\end{multline}
where we have used $d_L^{-1} \propto \sqrt{h_+^2 + h_\times^2} \sim \sqrt{(1+\cos^2\iota)^2 - 4 \cos^2\iota}$.
The effect here is to modify the observed distance or redshift distribution of
sources from the true distribution.  Since the effect enters at linear order in
$\kappa d_C$, it is degenerate with a variation in the BBH merger rate with
redshift; this is in contrast to effects which modify the leading-order relation
between the merger rate and distance or redshift, such as extra spacetime
dimensions \citep{Fishbach:2018edt,Pardo:2018ipy}.  The latter are, in
principle, observable even in a nearby sample of BBH mergers, with $z \to 0$. In this study, we use the values for $d_C$ reported in GWTC-2, without considering
these higher-order corrections.

Nevertheless, a full analysis could fit an evolving merger rate and
birefringence effects on inclination and amplitude, incorporating selection
effects.  Given the existing uncertainty about the evolution of the merger rate
with redshift \citep{Fishbach:2018edt,LIGOScientific:2018jsj} and the difficulty
in measuring $\cos \io$ with existing data (typical uncertanties are $\sim 0.3$
\citep{LIGOScientific:2018mvr}), our approximate analysis captures the majority
of the information about birefringence in the data at this time.

Note that in this study we assume that amplitude birefringence is the \textit{only} phenomenon that modifies the observed inclination angle from its true value. In particular, we do expect strong gravitational lensing to affect fewer than $10^{-3}$ of the detected events~\cite{Dai:2020tpj, Smith:2017mqu}, and hence do not consider strong lensing effects in this study.

Let us also discuss the effects of \textit{binary black hole precession}. In a generic BBH system in GR, spin-orbit coupling leads to a precession of the orbital place of the binary, which gives a time dependence to the inclination angle, $\iota(t)$, varying on the \textit{precession timescale}, which is longer than the orbital timescale. As discussed in~\cite{Alexander:2007kv}, precession has a different dependence on the instantaneous wavenumber at the detector than birefringence, so in principle it is possible to distinguish between precession and birefringence effects. While precession changes the ratio of left versus right circularly polarized GW radiation with inspiral time, GW amplitude birefringence will still preferentially amplify one or the other during propagation. Even if a detected event has a time-dependent $\cos \io (t)$, so long as $\cos \io(t) < 0$ or $\cos \io(t) > 0$ for all time, the event will be informative in measuring or constraining GW amplitude birefringence. Thus, precession is a systematic effect, but does not affect our ability to constrain amplitude birefringence.

In LIGO and Virgo, there are often too few cycles of BBH inspiral to significantly detect precession as measured by $\chi_p$, the combination of the BH spin components in the orbital plane~\cite{Abbott:2020niy} (note that the gravitational waveform models used in the GWTC-2 analysis, including NRSur7dq4, do include precession). In Eq.~\eqref{eq:GRPolarization}, thus, we set $\iota$ to a constant, and we will verify this assumption by checking whether the events in GWTC-2 that have a preferred $\cos \io$ (and hence are the most informative) have evidence of precession as reported by the GWTC-2 analysis.

Finally, in Eq.~\eqref{eq:DeltaProduct}, when combining all of the gravitational wave events to get an overall likelihood distribution on $\Delta$, we weigh each event equally, without considering signal to noise ratio (SNR), for example. Including such a statistic would be difficult, as, though events that have a smaller value of $d_C$ typically have higher SNRs, the resulting amplitude birefringence effect will be lower, due to signal polarizations having less cosmological distance over which to be enhanced and suppressed. Thus, we do not include additional weighting factors in Eq.~\eqref{eq:DeltaProduct}.

\section{GWTC-2 constraints on amplitude birefringence}
\label{sec:observed_constraint}

\begin{figure}
  \includegraphics[width=\columnwidth]{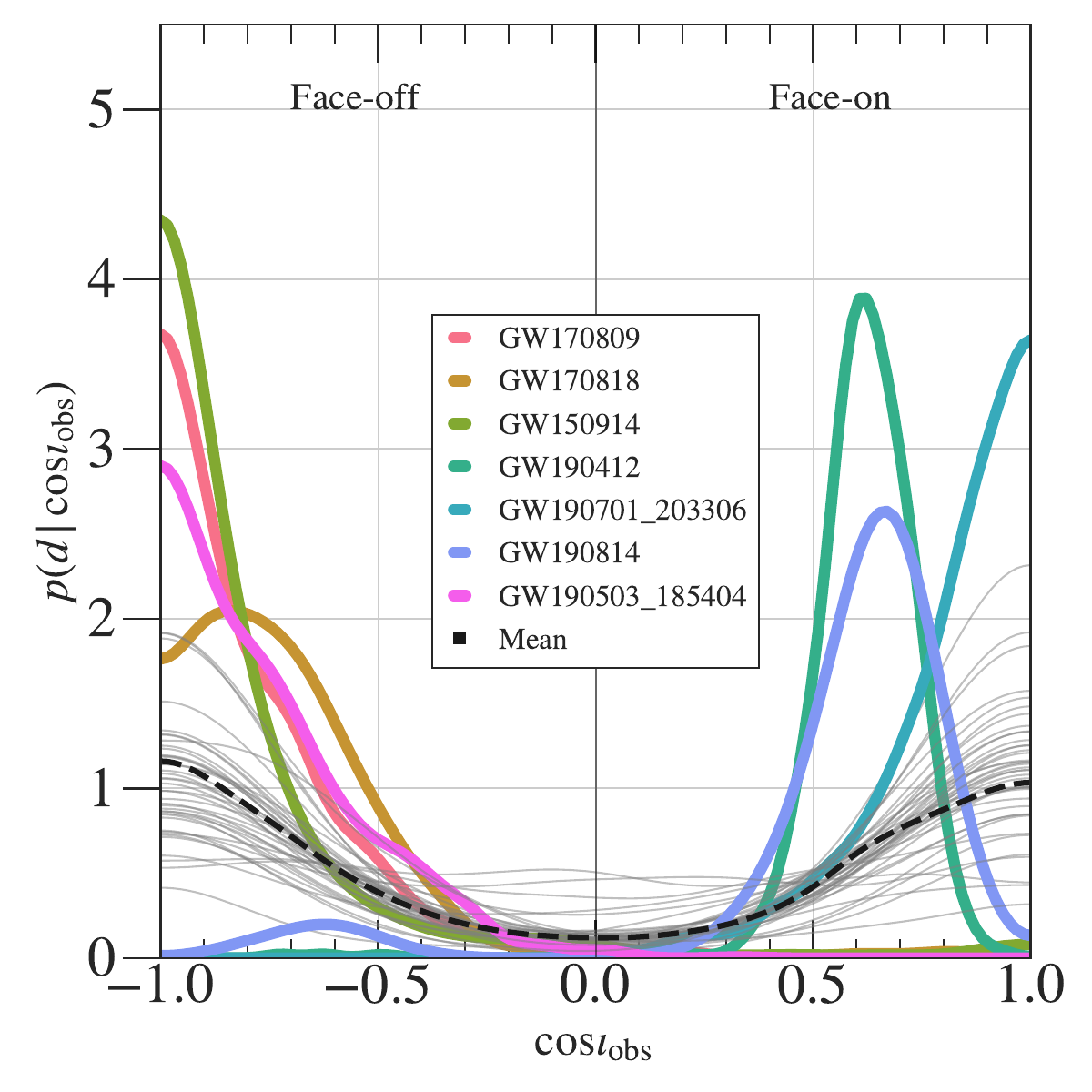}
  \caption{
  Likelihood distributions on $\cos \io$, the \textit{observed} inclination angle from GWTC-2~\cite{LIGOScientific:2018mvr,GWOSC:PE, Abbott:2020niy}. Each solid curve (including the gray curves) corresponds to a BBH detection, and the dashed black curve corresponds to the mean of $\cos \io$ across these events, weighing all events equally. While most events do not provide a confident measurement of $\cos \io$, we have highlighted (in thick, colored lines) the events that do show a strong preference for being face-off or face-on. Note that a population consistent with GR will have a mean distribution for $\cos \iota$ symmetric about zero.
  }
  \label{fig:CosIotaPosterior}
\end{figure}

In Fig.~\ref{fig:CosIotaPosterior}, we show the posterior distributions on the \textit{observed} inclination angle, $\cos \io$, from GWTC-2~\cite{LIGOScientific:2018mvr,GWOSC:PE, Abbott:2020niy}.\footnote{When available, we use the NRSur7dq4 parameter estimation results. Otherwise, if available, we use the SEOBNRv4PHM results, and finally we otherwise use the SEOBNRv4P results. We estimate that any systematic difference between which waveform model we use is well below the uncertainty in $\cos \iota$.}

The first two Advanced LIGO and Virgo observing runs, O1 and O2, contain 10 significant BBH detections, three of which have an inclination constraint sufficient to confidently identify the handedness of the wave, with each preferring a left-handed polarization (i.e.\ come from a binary orbiting in a left-handed sense with respect to the line-of-sight). The O3a observing run, meanwhile, contains approximately 37 candidate BBH detections, four of which provide a sufficient inclination constraint, with one left-handed polarization event, and three right-handed polarization events. While this results in a total of seven \emph{confident} inclination angle measurements, we will consider all of the $\cos \io$ distributions in our analysis, incorporating even weak preferences for left or right handed orbits from each one into our analysis.

Note that in the presence of strong amplitude birefringence, we would expect to observe such events with only one inclination angle preference. Thus, GWTC-2 rules out the possibility of \textit{purely} right or left-handed gravitational events. Due to their relative proximities and the thus correspondingly weak expected opacity constraints, we simplify our analysis by excluding the binary neutron star events. Thus, we exclude GW170817 and GW190425, as well as the neutron star - black hole candidate GW190426\_152155. Note that we do include GW190814, which provides a strong inclination constraint, but does have an (uncategorized) component mass of $2.59 M_\odot$~\cite{Abbott:2020niy}.

As discussed in Sec.~\ref{sec:additional_considerations}, we have assumed that $\io$ is constant, assuming that the GW events do not include precession of the orbital place. We can verify this assumption by considering the evidence of precession reported in GWTC-2 for the informative events highlighted in Fig.~\ref{fig:CosIotaPosterior}. Of these events, only GW190412~\cite{LIGOScientific:2020stg} confidently contains a non-zero spin component that is normal to the orbital angular momentum (cf. Fig. 11 in~\cite{Abbott:2020niy}). However, any precession in this system is ``marginal" (cf. Fig. 6 of~\cite{LIGOScientific:2020stg}), and hence we do not discard it from our sample.

\begin{figure}
  \includegraphics[width=\columnwidth]{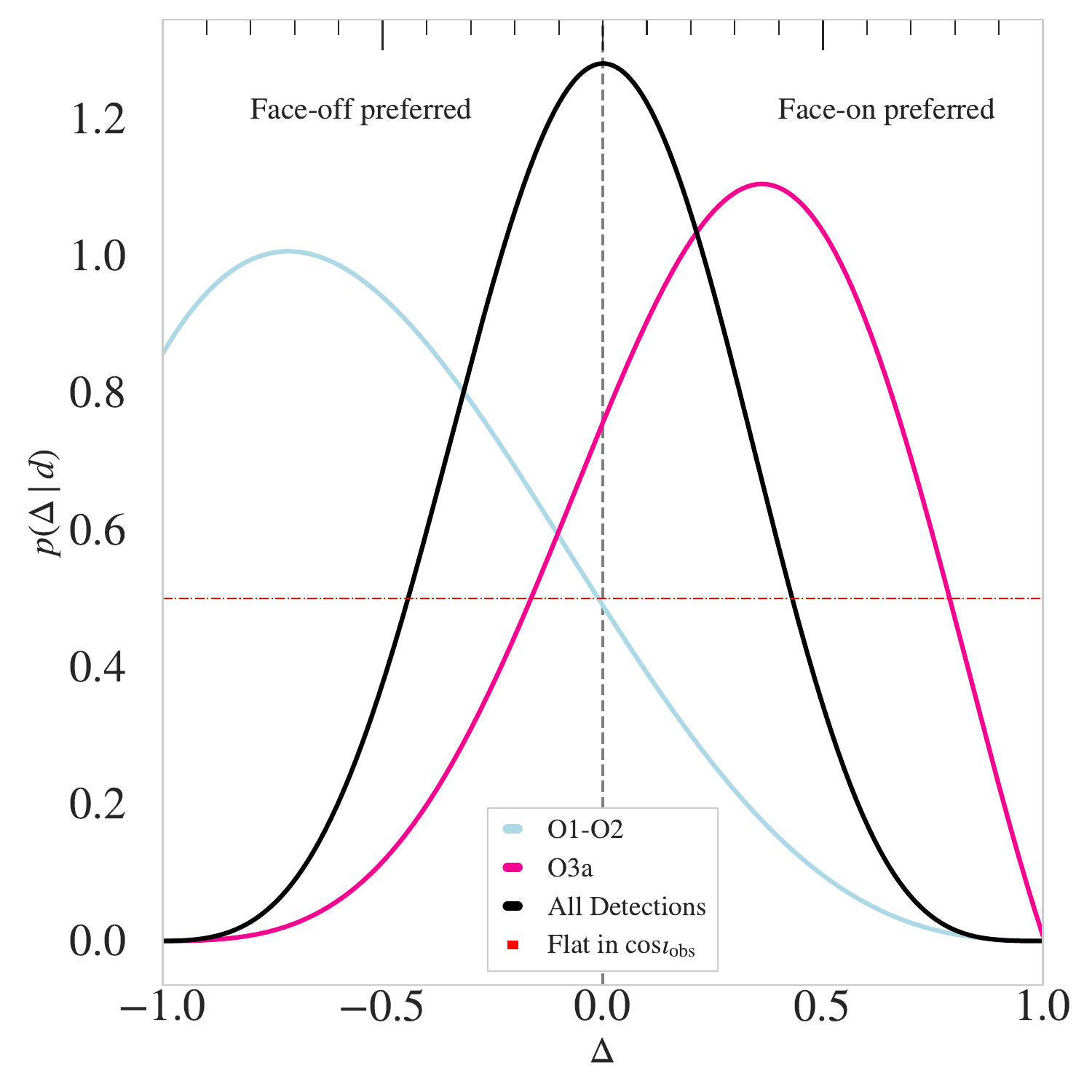}
  \caption{Posterior distribution for $\Delta$, which measures preference for face-on versus face-off observed events, as defined in Eq.~\eqref{eq:DeltaDef}. Without amplitude birefringence, the distribution should be symmetric around $\Delta = 0$. We show the distribution for $\Delta$ from O1-O2 events (light blue curve), and for O3a events (pink curve). We see that O1-O2 have a preference for face-off events, while O3a has a preference for face-on events. The resulting distribution is consistent with $\Delta = 0$, with a standard deviation of $0.4$, supporting no amplitude birefringence. The red dashed line, meanwhile corresponds to the values of $\Delta$ obtained by drawing from a distribution uniform in $\cos \io$ (thus corresponding to no information).
  }
  \label{fig:p-delta}
\end{figure}

Using these measures of $\cos \io$, we then compute a distribution on $\Delta$ from these observations using Eq.~\eqref{eq:DeltaProduct}, which we show in Fig.~\ref{fig:p-delta}. Note that to compute a posterior, $p(\Delta \mid d)$, from this likelihood, $p(d \mid \Delta)$, we must introduce a prior on $\Delta$, which we choose to be flat in $-1 < \Delta < 1$, given that we have no \textit{prior information} about $\Delta$. We see that the distribution on $\Delta$ from the O1-O2 observing runs disfavors face-on events, while preferring face-off events, and that the distribution on $\Delta$ from O3a disfavors face-off events, while preferring face-on events. Together, all of the detections are consistent with $\Delta = 0 \pm 0.4$ consistent with no amplitude birefringence.

Given $\Delta$, we can now use Eq.~\eqref{eq:DeltaKappa} to obtain a distribution on the absolute values of the opacity parameter $\kappa$, defined in Eq.~\eqref{eq:hKappa}. This will provide a physical measure of the amount of amplitude birefringence, the magnitude of which can then be used to constrain various beyond-GR theories. As detailed in Sec.~\ref{sec:BirefringenceMeasurement}, for our constraints on $\kappa$ and our projections, we use a common comoving distance to our BBH mergers of $d_C = d_C\left( z = 0.3 \right) \simeq 1.23 \, \mathrm{Gpc}$, corresponding to the median detected redshift in GWTC-2. We will additionally consider an analysis with $d_C = d_C(z = 0.3 \pm 0.1)$ in order to provide some error region for our results.

\begin{figure}
  \includegraphics[width=\columnwidth]{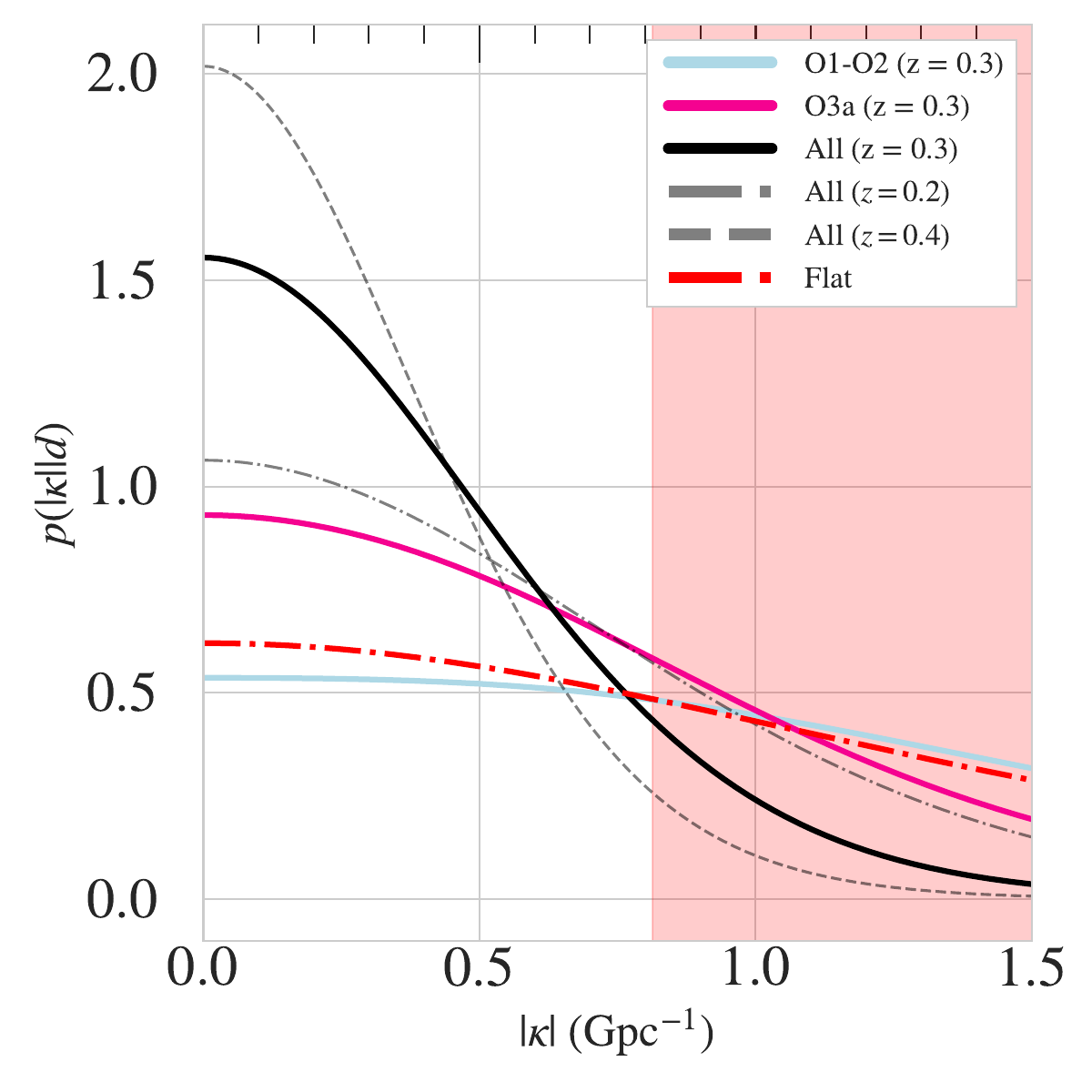}
  \caption{
  Distribution for the norm of the opacity parameter $\kappa$, as given in Eq.~\eqref{eq:hKappa}, which measures the strength of the amplitude birefringence effect. In the absence of amplitude birefringence, we expect $\kappa = 0$. Here, we compute the posterior on $|\kappa|$ with O1-O2 (light blue curve), and O3a (pink curve), combining all of the detections in the black curve. In order to show the effect of our assumption of a common comoving distance of $d_C(z = 0.3)$ for all events, we also plot lines (light gray), for $d_C(z = 0.3 \pm 0.1)$. The shaded region corresponds to $\kappa = 1/d_C(z = 0.3)$, in which the effective field theory assumption that $|\kappa| d_C \ll 1$ does not hold. The O1-O2 result on its own is uninformative, as it qualitatively agrees with a constraint generated from a flat, uninformative distribution in $\cos \io$ (dashed thick line). Adding in the O3a results, however, does result in an informative constraint.}
  \label{fig:KappaLikelihood}
\end{figure}

We show the resulting distribution on $\kappa$ in Fig.~\ref{fig:KappaLikelihood}. We observe that for a common comoving distance of $d_C = d_C(z = 0.3)$ (median detected redshift in GWTC-2), we can bound, at $1\sigma$:
\begin{align}
    \textrm{O1-O2:}\,\, &\kappa \lesssim 2.0 \textrm{ Gpc}^{-1} \,,\\
    \textrm{O3a:}\,\, &\kappa \lesssim 1.3 \textrm{ Gpc}^{-1}\,, \\
    \textrm{All:}\,\, &\kappa \lesssim 0.74 \textrm{ Gpc}^{-1} \,.
\end{align}
In Fig.~\ref{fig:KappaLikelihood}, we also show results for $\kappa$ for common comoving distances of $d_C = d_C(z = 0.3 \pm 0.1)$ for all of the detections, in order to qualitatively show the effect of a spread in the distance measurements on the inferred value of $\kappa$. These differences of $z \pm 0.1$ shift the inferred value for all of the detections by $\pm 0.25 \textrm{ Gpc}^{-1}$.

Recall that for the effective field theory to be valid, we require that $\kappa d_C \ll 1$. The analysis presented in this paper in terms of the observed inclination angle works for any value of $\kappa$, but we must be careful about the distances $d_C$. Thus, in Fig.~\ref{fig:KappaLikelihood} we shade the region for which $\kappa d_C > 1$, where this condition is violated given our choice of $d_C = d_C(z = 0.3)$.

In order to see how much information we have gained from these detections, let us consider a distribution flat in $\cos \io$ (meaning that all measured inclination angles are equally likely and $\cos \io$ carries no information about the system). The posterior on $\Delta$ for 47 events from this distribution using Eq.~\eqref{eq:DeltaDef} should be uniform on $\Delta$ (the events carry no information about which handedness is preferred). For such uninformative measurements if we wish to recover the correct flat distribution for $\Delta$ from our computations, we must satisfy the criterion that the number of samples used for each event is much larger than the number of detections as detailed in Appendix~\ref{sec:uninformative_appendix}.

If we then compute $\kappa$ from these values of $\Delta$ in Fig.~\ref{fig:KappaLikelihood}, we obtain a distribution that looks like that of O1-O2. We can thus conclude that the measurements of $\cos \io$ in O1-O2 are not sufficient to provide an informative constraint on $\kappa$; almost all of our constraint on $\kappa$ comes from the assumed prior on $\Delta$ transformed through Eq.\ \eqref{eq:DeltaKappa} into a prior on $\kappa$. However, adding in O3a does make the result deviate from the prior, thus showing that we can indeed constrain the level of amplitude birefringence with all of the BBH detections. 

In order to quantify this information, we can compute the Jensen-Shannon (JS) divergence $D_\mathrm{JS} (p (\lambda) \mid q (\lambda))$ of a distribution $p$ with respect to $q$. While technical details can be found in~\cite{Abbott:2020niy}, the KL divergence is a distance measure in units of bits of how a probability distribution is different from a reference probability distribution, thus allowing us to compare the curves in Fig.~\ref{fig:KappaLikelihood}. The JS divergence is a smoothed and symmetrized version of the Kullback-Leibler (KL) divergence~\cite{kullback1951}, and is particularly useful because it is guaranteed to be $0 \leq D_\mathrm{JS} \leq 1$ bit. The KL divergence is defined as

\begin{align}
\label{eq:KLDivergence}
    D_\mathrm{KL} (p (\lambda) \mid q (\lambda)) \equiv \int p (\lambda) \log_2 \left[ \frac{p (\lambda)} {q (\lambda)} \right] \,,
    d\lambda
    \,.
\end{align}
and the JS divergence is further defined as 
\begin{align}
\label{eq:JSDivergence}
     D_\mathrm{JS}(p, q) \equiv \frac{1}{2} \left(D_\mathrm{KL}(p \mid s) + D_\mathrm{KL}(q \mid s) \right) \,,
\end{align}
where $s = (p + q)/2$ is the average distribution.

Using the flat distribution as our reference distribution to compute the JS divergences for the distributions in Fig.~\ref{fig:KappaLikelihood}, finding
\begin{align}
 D_\mathrm{JS}(P_\textrm{O1-O2} (\kappa) \mid  P_\textrm{Flat} (\kappa)) &= 4.9 \times 10^{-4} \,, \\   
 D_\mathrm{JS}(P_\textrm{O3a} (\kappa) \mid P_\textrm{Flat} (\kappa)) &= 1.7 \times 10^{-2} \,, 
 \\ D_\mathrm{JS}(P_\textrm{All} (\kappa) \mid P_\textrm{Flat} (\kappa)) &= 7.1 \times 10^{-2}\,,
\end{align}
in units of bits. 

In order to interpret these quantities, we can compare to the values of $ D_\mathrm{JS}$ considered \textit{statistically significant} in the LIGO literature. For example, \texttt{Bilby}, a GW data analysis package, considers JS values greater that $2.9 \times 10^{-3}$ bits to be statistically significant~\cite{Romero-Shaw:2020owr}. Meanwhile, when considering precession in BBH systems, the LIGO GWTC-2 study considered events with JS values greater than $5 \times 10^{-2}$ bits as significant~\cite{Abbott:2020niy}. Thus, we conclude that the GWTC-2 results for $\kappa$, with $ D_\mathrm{JS} = 7.1 \times 10^{-2}$, are statistically significant.

% \begin{align}
%  D_\mathrm{KL}(P_\textrm{O1-O2} (\kappa)\, \|\,  P_\textrm{Flat} (\kappa)) &= 2.7 \times 10^{-3} \,, \\   D_\mathrm{KL}(P_\textrm{O3a} (\kappa)\, \|\,  P_\textrm{Flat} (\kappa)) &= 6.5 \times 10^{-2} \,, \\ D_\mathrm{KL}(P_\textrm{All} (\kappa)\, \|\,  P_\textrm{Flat} (\kappa)) &= 3.8 \times 10^{-1}\,,
% \end{align}
% in units of bits.

\section{Constraints on Chern-Simons gravity}
\label{sec:cs_implications}

We now use the inferred opacity $\kappa$ from Sec.~\ref{sec:observed_constraint} to place constraints on Chern-Simons gravity (CS). CS modifies the Einstein-Hilbert action of GR through the inclusion of a scalar field coupled to a term quadratic in spacetime curvature. In CS, amplitudes of left versus right circularly-polarized modes are exponentially enhanced and suppressed during propagation, with the strength of this \textit{amplitude birefringence} being governed by properties of the CS scalar field~\cite{Alexander:2009tp}. Thus, by placing constraints on the opacity parameter with GWTC-2, we can place observational constraints on CS.

Following the conventions of~\cite{Alexander:2009tp}, the action of Chern-Simons gravity takes the form
\begin{align}
\label{eq:Action}
    S = \int d^4 x \sqrt{-g} \Big( \frac{R}{16 \pi G} + \frac{1}{4} \alpha \vartheta \pont - \beta \frac{1}{2}\nabla_a \vartheta \nabla^a \vartheta \Big)\,,
\end{align}
where $g_{ab}$ is the spacetime metric with covariant derivative $\nabla_a$. The first term corresponds to the Einstein-Hilbert action of GR, where $R$ is the spacetime Ricci scalar. The second term couples the CS scalar field $\vartheta$ to spacetime curvature via the Pontryagin density $\pont \equiv \dual R_{abcd}R^{abcd}$, which is the spacetime Riemann tensor contracted with its dual~\cite{Alexander:2009tp}. The last term is a kinetic term for the scalar field, with constant $\beta$. We follow the choice of~\cite{Jackiw:2003pm, Alexander:2007kv}, and set $\alpha = (16 \pi G)^{-1}$, which gives $\vartheta$ units of length squared, $[\vartheta] = L^2$.

In \textit{non-dynamical} CS gravity, we set $\beta = 0$, and $\vartheta$ is `frozen-in' with some pre-defined profile~\cite{Jackiw:2003pm}, which we will leave unspecified for now. Note that $\vartheta$ cannot be constant, otherwise the $\pont$ term, a topological invariant, would integrate out of the action in Eq.~\eqref{eq:Action}. The resulting theory, however, is not diffeomorphism invariant. However, provided that $\pd_t \vartheta$ is small enough, we can treat this as a cosmological solution of \textit{dynamical} Chern-Simons gravity~\cite{Alexander:2009tp}.

As calculated by Alexander et al.~\cite{Alexander:2007kv}, in CS, GWs propagating through a Friedmann-Lema\^{i}tre-Robertson-Walker universe are  exponentially suppressed and enhanced depending on helicity. For compact-binary sources, this \textit{birefringence} effect manifests in a change in the observed inclination of the binary, $\cos \io$, from the true inclination angle of the source, $\cos \iota$, as
\begin{align}
\label{eq:hZeta}
    \nn \left(\frac{h_\Ro}{h_\Lo}\right)_\mathrm{CS} &= \left(\frac{1 + \cos \iota}{1 - \cos \iota}\right)^2 \exp \left[ \frac{2 k(t)}{H_0} \zeta (\vartheta) \right] \\
    &= \left(\frac{1 + \cos \io}{1 - \cos \io}\right)^2\,.
\end{align}
Here, we have used the conventions of~\cite{Alexander:2007kv, Alexander:2009tp}, where $k(t)$ is the wavenumber for the given Fourier propagating mode, with units of $L^{-1}$, and $\zeta(\eta)$ is a dimensionless function of the integrated history of the CS scalar field. While Eq.~\eqref{eq:hZeta} is a function of the wavenumber, we will estimate that $k(t)$ covers a narrow frequency range, and thus write $k(t) \sim k$, where $k$ is a typical value in this range, without treating each mode separately.

In~\cite{Alexander:2007kv}, the authors calculate $\zeta(\eta)$, a dimensionless function of the integrated history of the CS scalar-field, for a matter-dominated universe (with scale factor $a(\eta) = a_0 \eta^2$, where $a_0$ is the present-day value and $\eta$ is conformal time). Since the LIGO sources are found at redshifts $z < 1$ (300--3000 Mpc)~\cite{LIGOScientific:2018mvr}, we focus on a dark-energy dominated universe, with $a(t) = a_0 e^{H_0 t}$. We compute the corresponding $\zeta$, in terms of dimensionless conformal time $\eta$, to be
\begin{align}
\label{eq:DarkEnergyZeta}
    \zeta(\eta) = \frac{H_0^2}{2} \int_\eta^1 \left( \eta^2 \vartheta''(\eta)  - 2 \eta \vartheta'(\eta) \right) d \eta\,.
\end{align}
We give the full calculation in Appendix~\ref{sec:zeta_appendix}.

\subsection{General constraint}
\label{sec:general_cs_constraint}

Comparing Eq.~\eqref{eq:hZeta} with Eq.~\eqref{eq:hKappa}, we can directly relate $\zeta(\eta)$, which captures all of the dependence on the CS field, to the measured value of $\kappa$ as
\begin{align}
\label{eq:ZetaKappa}
    \zeta(\eta) = \frac{\kappa d_C H_0}{k} \,.
\end{align}
Thus, setting $d_C\left( z = 0.3 \right) \simeq 1.23$ Gpc for a typical Advanced LIGO BBH source distance (corresponding to the median detected redshift in GWTC-2)~\cite{LIGOScientific:2018mvr}, and setting $k \sim 2 \pi \times 100 \,\mathrm{Hz} / c \sim 2 \times 10^{-6} \, \mathrm{m}$ for the approximate value of the region of greatest sensitivity of LIGO (cf.~\cite{TheLIGOScientific:2016agk, LIGOScientific:2018mvr}), we obtain the dimensionless result
\begin{align}
\label{eq:ZetaValue}
    \zeta(\eta) = \left(\frac{\kappa}{1 \textrm{ Gpc}^{-1}} \right) \times 6.6 \times 10^{-21}\,.
\end{align}
From the results for GWTC-2 in Sec.~\ref{sec:observed_constraint}, we compute
\begin{align}
     \textrm{O1-O2:}\,\, &\zeta(\eta) \lesssim 1.3 \times 10^{-20} \,,\\
     \textrm{O3a:}\,\, &\zeta(\eta) \lesssim 8.6 \times 10^{-21}\,, \\
     \textrm{All:}\,\, &\zeta(\eta) \lesssim 4.9 \times 10^{-21} \,.
\end{align}

In the above expressions, we have left the `frozen-in' profile of $\vartheta$ unspecified. Let us suppose that $\vartheta$ is dependent on some CS parameter $P$. For some specified profile $\vartheta[P]$, the reader can thus use Eqs.~\eqref{eq:ZetaValue} and~\eqref{eq:DarkEnergyZeta} to compute a value of $P$ given a value of $\kappa$.

\subsection{Constraint on canonical $\vartheta$ profile}
\label{sec:canonical_cs_constraint}

Let us now consider the `canonical' profile for $\vartheta$ given in~\cite{Jackiw:2003pm, Alexander:2009tp, Yunes:2008ua}, where $\vartheta$ has an isotropic, time-dependent profile of the form
\begin{align}
\label{eq:ThetaPi}
    \vartheta = \frac{t}{\mu}\,,
\end{align}
where $\mu$ is a mass scale with units $[\mu] = L^{-1}$. Note that when $\mu$ is large, we recover GR.

\begin{figure}
  \includegraphics[width=\columnwidth]{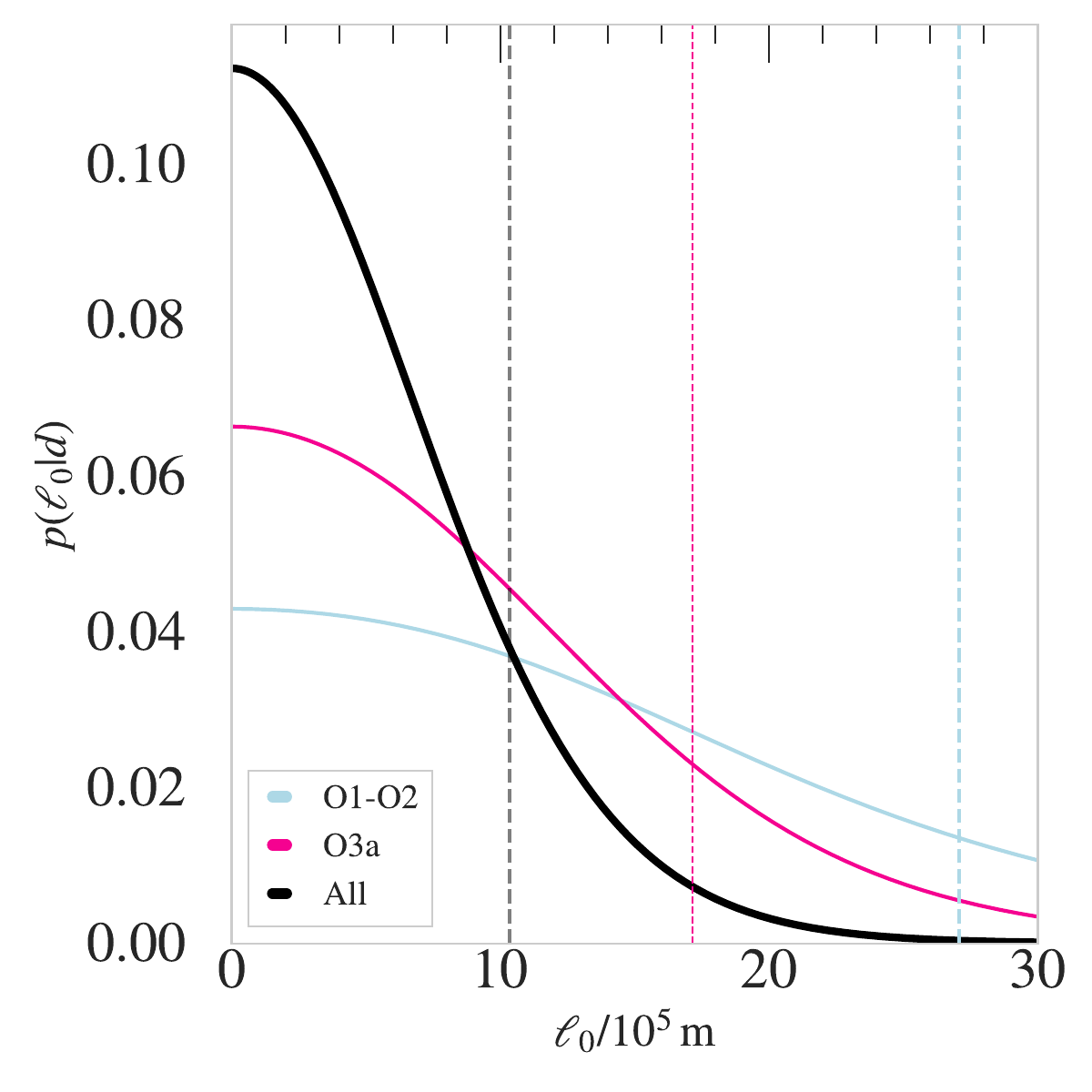}
  \caption{Posterior on $\ell_0$, the CS field length scale for the canonical CS field profile given in Eqs.~\eqref{eq:ThetaPi} and~\eqref{eq:L0Definition}. We compute the likelihood from the observations in O1-O2 (light blue curve), O3a (pink curve), and both catalogs (black curve). Each vertical line corresponds to $1-\sigma$.
  }
  \label{fig:L0Likelihood}
\end{figure}

Let us define
\begin{align}
\label{eq:L0Definition}
    \ell_0 \equiv \frac{1}{\mu}
\end{align}
to be the CS lengthscale for this field profile. With this profile, $\zeta(\eta)$ in Eq.~\eqref{eq:DarkEnergyZeta} becomes
\begin{align}
\label{eq:ZetaPi}
    \zeta(\eta) = \frac{3 H_0 \ell_0}{2 c} (1 - \eta) = \frac{3 H_0 \ell_0 d_C}{2 d_H}\,.
\end{align}
where we have re-introduced a factor of c and have set $(1 - \eta) \sim d_C / d_H$, where $d_H \equiv c/H_0$ is the Hubble distance. Combining Eqs.~\eqref{eq:ZetaKappa} and~\eqref{eq:ZetaPi}, we obtain
\begin{align}
    \ell_0 = \frac{2 c d_H \kappa }{3 k}\,.
\end{align}
which becomes
\begin{align}
\label{eq:EllPi}
    \ell_0 = \left(\frac{\kappa}{1 \textrm{ Gpc}^{-1}} \right) \times 1400 \textrm{ km}\,.
\end{align}
Given the posterior on $\kappa$ computed in Sec.~\ref{sec:observed_constraint}, we show the posterior on $\ell_0$, computed using Eq.~\eqref{eq:EllPi} in Fig.~\ref{fig:L0Likelihood}. We can thus bound

\begin{align}
     \textrm{O1-O2:}\,\, &\ell_0 \lesssim 2.8 \times 10^3  \textrm{ km} \,,\\
     \textrm{O3a:}\,\, &\ell_0 \lesssim 1.8 \times 10^3  \textrm{ km}\,, \\
     \textrm{All:}\,\, &\ell_0 \lesssim 1.0 \times 10^3 \textrm{ km} \,.
\end{align}

Note that while we have assumed the `canonical' profile for $\vartheta$, this result is also a good approximation if the second time derivative of $\vartheta$ is small, meaning that for a small-enough time, the field profile is linear in time.

\subsection{Projected value of $\ell_0$ with more detections}
\label{sec:projected}

We can project future constrains on $\ell_0$ using the difference in constraints we have obtained with GWTC-2 results to see how this constraint would improve with future detections. At fixed detector sensitivity, we expect that the constraint will go as $\ell_0(N)  \sim 1 / \sqrt{N}$, where $N$ is the number of detections. But as the detector sensitivity changes so does the typical distance to a detected merger.  Since advanced LIGO at design sensitivity is expected to have a larger reach in distance~\cite{Aasi:2013wya}, we set the typical value of the redshift to $z = 0.75$. Repeating the previous analysis with the O1-O2 detections and the O3a detections, and with $z = 0.75$ instead of $z = 0.3$, we find that
\begin{align}
    \ell_0 (N = 10, z = 0.75) &= 1300 \textrm{ km} \\
    \ell_0 (N = 47, z = 0.75) &= 490 \textrm{ km}
\end{align}
With 1000 BBH detections at design sensitivity, for example, we would expect to bound $\ell_0 \lesssim 100 \, \mathrm{km}$. This projection is the result of two anticipated improvements---first in the greater reach in redshift of LIGO at design sensitivity, and second in the number of detections.

\subsection{Implications of Chern-Simons constraint}

Let us compare the physical constraint on the canonical Chern-Simons lengthscale from Sec.~\ref{sec:canonical_cs_constraint} to additional observed bounds on the non-dynamical theory, using the $1000$ km bound we obtain from GWTC-2. Smith et al.~\cite{Smith:2007jm}  used Solar-System measurements of frame-dragging from LAGEOS and Gravity Probe B to bound $|\dot \vartheta| \leq 3000 (\kappa/\alpha)$ km. We have chosen $\alpha = \kappa$ in this study, and for the canonical profile, we have $\dot \vartheta = \ell_0$. Hence, the Smith et al.~constraint becomes $\ell_0 \leq 3000$ km. The bound from GWTC-2 is smaller than this number, indicating that LIGO events can constrain the non-dynamical theory more tightly than this Solar-System test.

Alexander et al. proposed an amplitude birefringence analysis with LISA~\cite{Alexander:2007kv}, estimating that for a $10^6\,M_\odot$ BBH at redshift $z \sim 15$, one could bound $\ell_0 \leq 10^{-2}$ km~\cite{Alexander:2009tp}.\footnote{Note that the analysis in this paper was performed for a \textit{dark-energy} dominated universe, which is applicable to LIGO sources with $z \sim 1$, while the LISA analysis required a matter-dominated universe.} This is a stronger bound that the one obtained in this paper, and attempting to achieve such a bound with LIGO-Virgo events would require $N \sim 10^{11}$ detections (cf. Sec.~\ref{sec:projected}). The authors of~\cite{Alexander:2007kv} perform a Fisher-matrix analysis for a source sweeping through $10^{-4} - 10^{-2}$ Hz, keeping track of the frequency dependence in $k(t)$ and hence $\io(t)$. In this study, we have approximated $k(t)$ as a constant $2\pi \times 100$ Hz, which in turn corresponds to setting $\io(t)$ to a constant function of time-varying apparent inclination angle described in~\cite{Alexander:2007kv}. While LISA is sensitive to this effect due to probing long BBH inspirals, LIGO is not sensitive to this effect, as there are not enough cycles in the LIGO band to probe precession for most events~\cite{LIGOScientific:2018mvr}.

Additionally, Hu et al.~\cite{Hu:2020rub} performed a study analyzing the capability of a network of future space-based detectors (LISA, Taiji, and TianQin) to constrain parity violations in gravitational wave propagation, finding that for a $10^6\,M_\odot$ event at 20 Gpc, the parity violating scale from \textit{amplitude birefringence} could be bounded to $M_\mathrm{PV} > \mathcal{O}(10^{-15})$ eV, corresponding to $2 \times 10^{5}$ km. This, as the authors note, is a weaker bound than the constraint from ground-based detectors.

Yunes and Spergel~\cite{Yunes:2008ua} performed a binary pulsar test with PSR J0737--3039, finding $\ell_0 \lesssim 6 \times 10^{-9}$ km, a bound much stronger than the one reported in this paper. The periastron precession of a system is corrected in CS, with the gradient of $\vartheta$ selecting a preferred direction in spacetime for the correction. The strength of this correction relative to GR is governed by $a^2/R^2$, where $a$ is the semimajor axis of the system, and $R$ is the radius of the object. With a large separation ($\sim 10^6$ stellar radii in this case), and small radii, a binary pulsar system produces a very strong constraint. However, as shown in~\cite{AliHaimoud:2011bk}, this analysis failed to account for several effects that lead to a suppression of the rate of periastron precession. In particular,~\cite{Yunes:2008ua} modeled PSR J0737--3039B as a point particle, rather than an extended body with radius $R_B$. If $R_B$ is larger than $2 \pi \ell_0$ (the CS wavelength), the average force per unit mass is suppressed by a factor of $\sim 15 (\ell_0 / R_B)^3$. Thus, in order to match the observed constraint on periastron precession, $\ell_0$ must be $\gtrsim R_B$. Indeed,~\cite{AliHaimoud:2011bk} computed a corrected constraint of $\ell_0 \lesssim 0.4$ km.

In addition,~\cite{Yunes:2008ua} probes a different physical regime than we probe in this paper. Yunes and Spergel assume the canonical, \textit{global} $\vartheta = \ell_0 t$ profile, but use a \textit{local} measurement to probe $\ell_0$. This involves assuming that the canonical profile, which has no spatial dependence, truly holds within our galaxy, and that there are no spatial density variations in the field near PSR J0737-3039. In this paper, however, we use an \textit{integrated} history of $\vartheta$, sampling its temporal evolution, all the way from redshift $z \sim 1$ to present day. Over such cosmological distances, choosing the smooth, isotropic profile $\vartheta = \ell_0 t$ may be justified, as any spatial effects can be  presumed to integrate out. Thus, our analysis differs from binary pulsar tests in that we have used a global measurement to constraint a global quantity, without making any local assumptions.

Recently, Wang et al.~\cite{Wang:2020cub} analyzed the presence of amplitude \textit{and velocity} birefringence in GWTC-1, the first catalog of LIGO and Virgo detections, finding no evidence of parity violation. Their methods are different from the ones presented in this paper, as they match GWTC-1 data against GW templates that include birefringence effects, rather than looking at an ensemble of inclination angles. The constraint on the parity violating energy scale found in~\cite{Wang:2020cub} is $M_\mathrm{PV} > 0.07 $ GeV, which corresponds to a lengthscale of $\hbar c / M_\mathrm{PV} \sim 10^{-18}$ km. However, this comes from velocity birefringence effects, as LIGO is more sensitive to phase, rather than amplitude, modifications. Indeed, the constraint from amplitude birefringence effects only is $M_\mathrm{PV} > 10^{-22}$ GeV which corresponds to $\sim 2000$ km. Similarly, Yamada et al.~\cite{Yamada:2020zvt} performed a parametrized tests of parity violation in gravitational wave propagation for GWTC-1, finding a minimum bound of $\ell_0 \leq 1422$ km for GW151226 for CS gravity. Our GWTC-2 result of $\ell_0 \leq 1000$ km improves on both of these results.

\section{Conclusion}
\label{sec:conclusion}

In this study, we have used GWTC-2~\cite{Abbott:2020niy, GWOSC:PE}, including events from the first three observation runs, to perform a new test of general relativity (GR). We have placed an observational bound on gravitational wave amplitude birefringence, which is absent in GR, but present in various beyond-GR theories. Namely, we have bounded the opacity parameter governing the strength of the amplitude birefringence to $\kappa \lesssim 0.74 \textrm{ Gpc}^{-1}$ (Sec.~\ref{sec:observed_constraint}).

This general opacity constraint can then be mapped onto any beyond-GR theory exhibiting amplitude birefringence (see~\cite{Zhao:2019xmm} for a review). We have focused on (non-dynamical) Chern-Simons gravity, a beyond-GR theory with motivations in string theory and loop quantum gravity (Sec.~\ref{sec:cs_implications}). We have used our results for $\kappa$ to bound $\zeta(\eta)$, a general CS parameter governing the CS scalar field, to $\zeta(\eta) \lesssim 4.9 \times 10^{-21}$. We then computed the constraint on the CS lengthscale of the canonical scalar field profile, to give $\ell_0 \lesssim 1.0 \times 10^3 $ km (Sec.~\ref{sec:canonical_cs_constraint}).

One of the main benefits of our analysis is that it is simple and fast (of order minutes on one CPU), and only requires looking at inclination angle posterior distributions for gravitational wave events, which are readily available from LIGO and Virgo catalogs, without performing an independent parameter estimation analysis. We plan to repeat this analysis with future LIGO and Virgo observations, obtaining an even tighter bound on this beyond-GR effect.

\section*{Acknowledgements}

 MO and WF are funded by the  Center for Computational Astrophysics at the Flatiron Institute, which is supported by the Simons Foundation.
 MI~is supported by NASA through the NASA Hubble Fellowship grant \#HST--HF2--51410.001--A awarded by the Space Telescope Science Institute, which is operated by the Association of Universities for Research in Astronomy, Inc., for NASA, under contract NAS5--26555.

\appendix

\section{Joint distance-$\kappa$ analysis}
\label{sec:appendix_distance}

Let us now consider dropping the assumption used in the main analysis of Sec.~\ref{sec:observed_constraint} that all of the observed GW events are at the same distance. This requires performing a joint analysis for $d_L$ and $\kappa$.

For ease of notation, let us write
\begin{align}
\label{eq:abbreviated_vars}
     c &\equiv \cos \iota \,, \\
    \nn c_o &\equiv \cos \io \,, \\
    \nn d_{Lo} &\equiv d_{L, \mathrm{obs}} \,.
\end{align}

\begin{figure}
  \includegraphics[width=0.6\columnwidth]{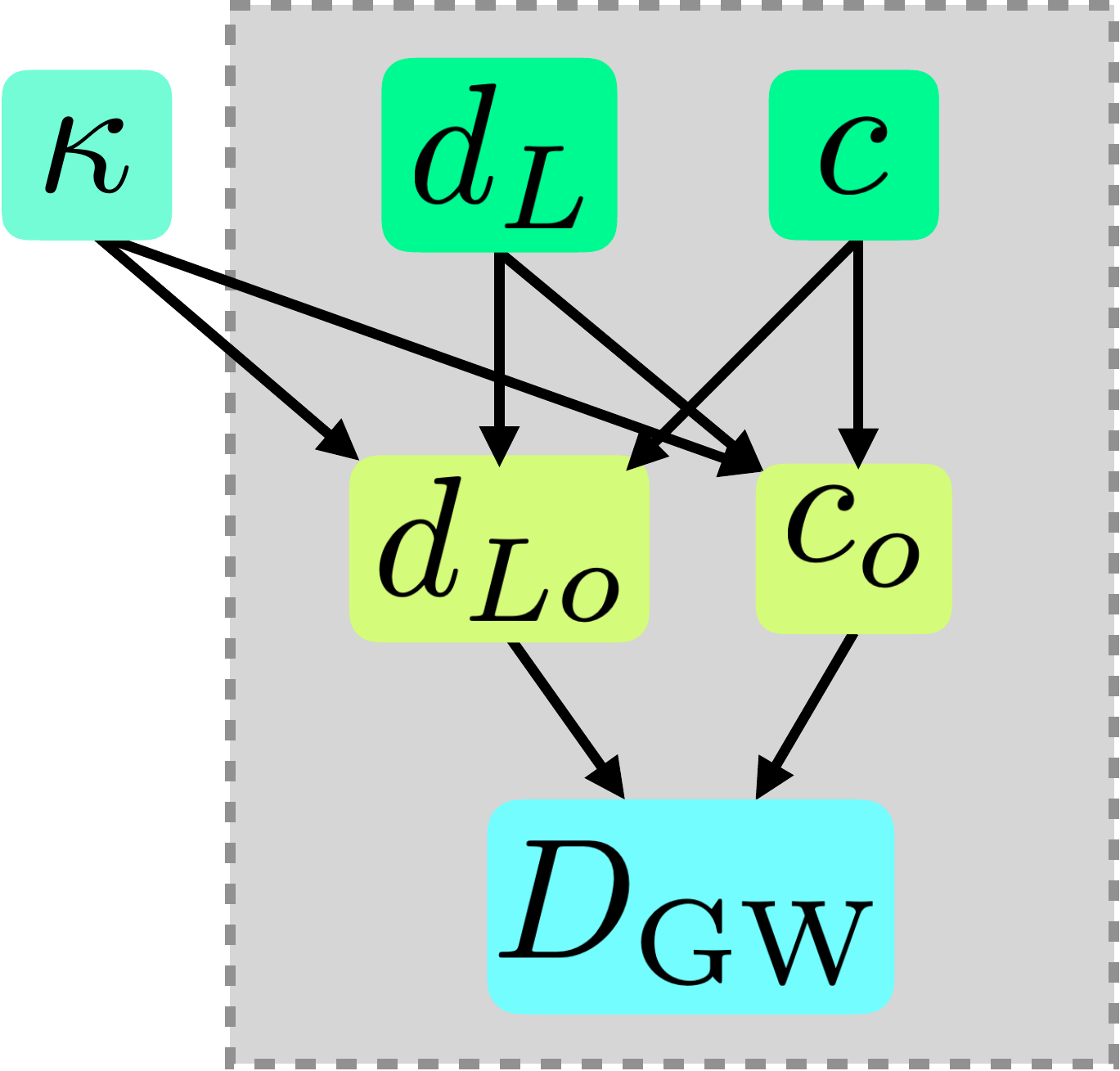}
  \caption{Probabilistic graphical model illustrating the relationship between the variables (see Eq.~\eqref{eq:abbreviated_vars} for abbreviations). The gray region represents data that applies to each gravitational wave event, while $\kappa$ is a universal constant, independent of each event. The true luminosity distance $d_L$, the true inclination angle $c$, and $\kappa$ affect the observed luminosity distance $d_{Lo}$ and the observed inclination angle $c_o$. These in turn affect the observed gravitational wave data $D_\mathrm{GW}$.
  }
  \label{fig:PGM}
\end{figure}

We can model this entire system as a probabilistic graphical model (PGM), as illustrated in Fig.~\ref{fig:PGM}. The directions in the PGM denote the \textit{influences} between various variables. In our case, $\kappa$, $d_L$, and $c$, which are true astrophysical parameters, influence the observed variables $d_{Lo}$ and $c_o$. In turn, $d_{Lo}$ and $c_o$ influence the observed gravitational wave data $D_\mathrm{GW}$. In this model, $\kappa$ plays a special role, because it is shared by the entire population.

For each event, we can marginalize over the distributions for the observed variables, $\{d_{Lo}, c_o\}$ to obtain $p(D_\mathrm{GW} \mid \kappa)$ as

\begin{align}
\label{eq:kappa_likelihood}
    p(D_\mathrm{GW} | \kappa) = \int \mathrm{d} c_o \, \mathrm{d} d_{Lo} \,p(D_\mathrm{GW} \mid c_o, d_{Lo}) p(c_o, d_{Lo} \mid \kappa) \,.
\end{align}
The above expression is a standard marginalization using the PGM, without making any astrophysical arguments.

We can compute the likelihood for $M_\mathrm{obs}$ gravitational wave observations as the product
\begin{align}
\label{eq:product}
p\left( \left\{ D_{\mathrm{GW},j} \mid j = 1, \ldots, M_\mathrm{obs} \right\} \mid \kappa \right) = \prod_{j=1}^{M_\mathrm{obs}} p\left( D_{\mathrm{GW},j} \mid \kappa \right)\,,
\end{align}
where we use Eq.~\eqref{eq:kappa_likelihood} to compute each of the individual likelihoods in the product.

Let us now work with Eq.~\eqref{eq:kappa_likelihood}, further marginalizing over $d_{Lo}$ as
\begin{align}
\label{eq:integral_kappa_likelihood}
    p(D_\mathrm{GW} \mid \kappa) &= \int \mathrm{d} c_o \, \mathrm{d} d_{Lo} \,p(D_\mathrm{GW} \mid c_o, d_{Lo}) \\
\nn    & \quad \times p(c_o  \mid \kappa, d_{Lo}) p(d_{Lo}) \,.
\end{align}

In order to compute $p(d_{Lo})$, we assert that the distribution of observed luminosity distances tracks the star formation rate, with
\begin{align}
\label{eq:p_d_Lo}
p\left( d_{Lo} \right) \propto \frac{\left( 1 + z \right)^{\alpha}}{1 + \left(\frac{1+z}{1+z_p} \right)^\beta} \frac{\mathrm{d} V}{\mathrm{d} z} \frac{\mathrm{d} z}{\mathrm{d} d_L} \frac{1}{1+z}
\end{align}
where $\alpha=2.7$, $z_p=1.9$, and $\beta=5.6$ from~\cite{doi:10.1146/annurev-astro-081811-125615}.  Effectively, we adjust the true merger rate evolution with redshift to match the observed distribution to the star formation rate (this is consistent with the population analysis in~\cite{LIGOScientific:2018jsj}).  We do this to avoid learning anything about $\kappa$ from any imposed prior on the \emph{true} merger rate evolution, since we are a priori very uncertain about it.

\begin{figure}
  \includegraphics[width=1.0\columnwidth]{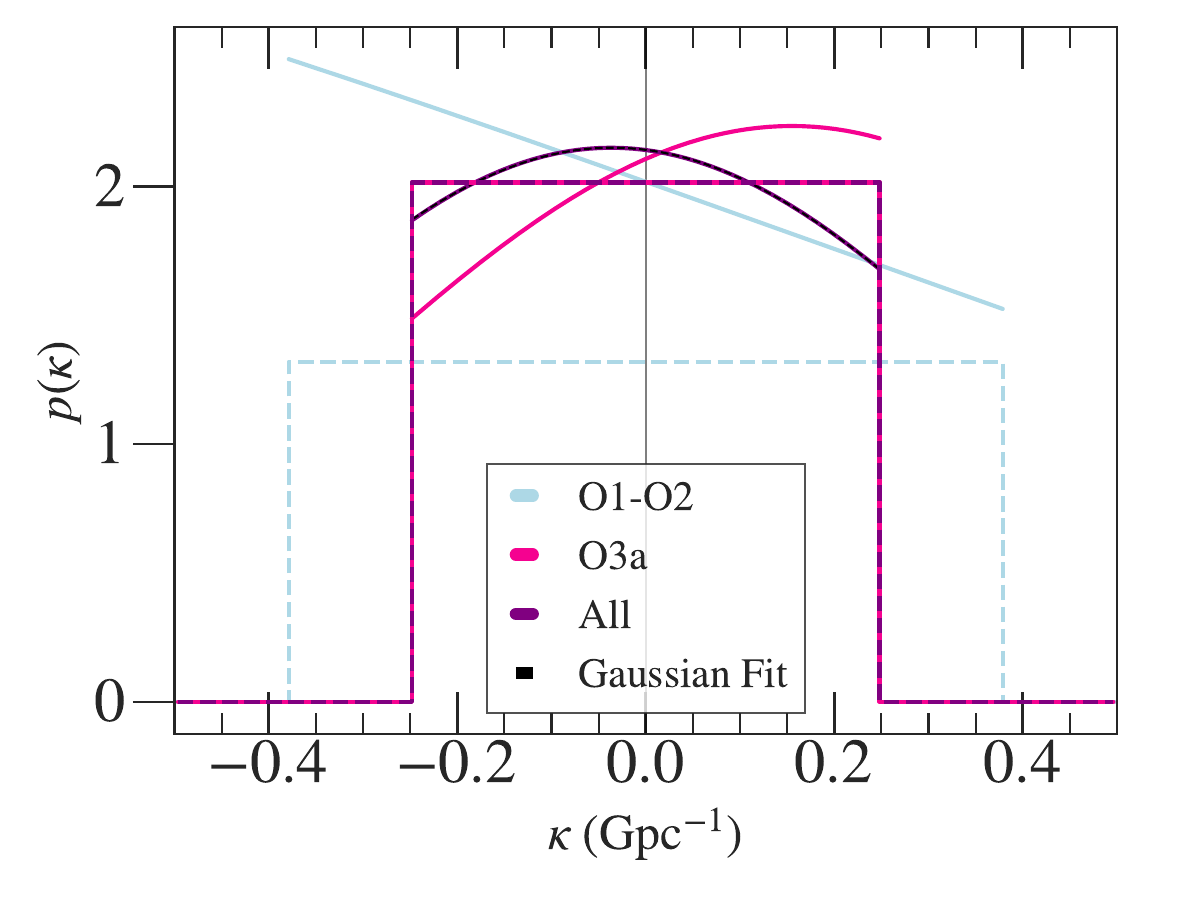}
  \caption{Posterior distribution on $\kappa$ using a joint distance-$\kappa$ analysis. We show the posteriors for the O1-O2 detections (light blue curve), O3a detections (pink curve), and all detections (purple curve). The combined result prefers $\kappa = 0$, thus showing consistency with GR. Compare to Fig.~\ref{fig:p-kappa_distance}, which assumes a fixed distance for all events. For each dataset, we show the corresponding prior on $\kappa$ with a dot-dashed line, given through the condition that $\kappa d_C \ll 1$. The priors are different for the two datasets as they have different maximum values of $d_C$. We also fit a Gaussian to all of the detections to estimate a variance for the distribution (black dashed curve, visually overlapping with the data).
  }
  \label{fig:p-kappa_distance}
\end{figure}

Now, we need to compute $p(c_o \mid \kappa, d_{Lo})$, We assume from isotropy that the true inclination angle at the source, $c$, is independent of $d_L$ and $\kappa$, giving
\begin{align}
    p(c \mid d_L, \kappa) = \frac{1}{2}\,.
\end{align}
Then, we can compute $p(c_o \mid \kappa, d_{Lo})$ through a substitution of variables as
\begin{align}
    p(c_o \mid \kappa, d_{Lo}) = p(c \mid d_L, \kappa) \left| \frac{\partial c}{\partial c_o} \right| \,.
\end{align}
From Eq.~\eqref{eq:Inclinations}, we can compute
\begin{align}
    \frac{\partial c}{\partial c_o} = 1 + c_o \kappa d_C + \mathcal{O} (\kappa d_C)^2\,.
\end{align}
Thus, we obtain
\begin{align}
    p(c_o \mid \kappa, d_{Lo}) = \frac{1}{2} \left(1 + c_o \kappa d_C + \mathcal{O} (\kappa d_C)^2 \right)\,.
\end{align}
Now we have all of the pieces of Eq.~\eqref{eq:integral_kappa_likelihood}. To make the above expressions valid, we impose that $\kappa d_C \ll 1$. We enforce this by choosing a flat prior on $\kappa$ symmetric about zero, with support up to maximum allowed value $\kappa_\mathrm{max}$ determined by the largest value of the distance, $d_{C\mathrm{, max}}$. We choose the 99th percentile value of $d_C$ in each dataset to give $d_{C\mathrm{, max}}$ (cf. Fig.~\ref{fig:p-kappa_distance} for an illustration).

In practice, we have access not to continuous probability distributions, but rather to $N$ samples from each gravitational wave events. Thus, we express the integral in Eq.~\eqref{eq:integral_kappa_likelihood} as a sum over $N$ samples, giving
\begin{align}
\label{eq:sum_kappa_likelihood}
    p\left( D_\mathrm{GW} \mid \kappa \right) &\simeq \frac{1}{N} \times \\
    \nn & \sum_{n=1}^{N} \frac{p\left(c_{o,n} \mid d_{Lo,n}, \kappa \right) p\left( d_{Lo,n} \right) p\left(\vec \theta_{n, \mathrm{other}} \right)}{p\left( \vec \theta_n \right)}\,.
\end{align}
The quantity, $\vec \theta_\mathrm{n}$ refers to all of the parameters of the model. The quantity $\vec \theta_\mathrm{n, other}$, meanwhile,  refers to all of the parameters besides the distances, inclination angles, and $\kappa$ in the model, such as the masses and spins of the black holes. We can use priors on $p(\vec \theta_\mathrm{n, other})$ to re-sample the distributions on parameters given in GWTC-2, with weights
\begin{align}
    w_n = \frac{p\left( d_{Lo,n} \right) p\left( \vec \theta_{\mathrm{other},n} \right)}{p\left( \vec \theta_n \right)}\,,
\end{align}
to give
\begin{align}
p\left( D_\mathrm{GW} \mid \kappa \right) \simeq \frac{1}{N'} \sum_{n=1}^{N'} w_n p\left(c_{o,n} \mid d_{Lo,n}, \kappa \right)\,,
\end{align}
for the sum in Eq.~\eqref{eq:sum_kappa_likelihood}.

In particular, in keeping with Eq.~\eqref{eq:p_d_Lo}, we want to choose the prior on masses and distances to track the star formation rate. The prior $p(m_1, m_2, d_{Lo})$ used in GWTC-2 is flat in detector frame masses and flat in $c_o$, of the form
\begin{align}
p\left( m_1, m_2, d_{Lo} \right) \propto \frac{\partial m_1^{\mathrm{det}}}{\partial m_1} \frac{\partial m_2^{\mathrm{det}}}{\partial m_2} d_L^2 = \left(1 + z\right)^2 d_L^2\,.
\end{align}
We will re-weight using a prior on the masses and is proportional to $m_1^{-1.6}$ and flat in mass ratio, $q$, the approximate best-fit distribution from~\cite{LIGOScientific:2018jsj}, of the form
\begin{align}
p\left( m_1, m_2, d_{Lo} \right) \propto m_1^{-1.6} \frac{\partial q}{\partial m_2} p\left( d_{Lo} \right) = m_1^{-2.6} p\left( d_{Lo} \right),
\end{align}
where $p\left( d_{Lo} \right)$ tracks the star formation rate as given in Eq.~\eqref{eq:p_d_Lo}. Note that we do not consider the parameter space of other physical binary black hole populations in this study, in part because population models are not presently well-constrained with GWTC-2~\cite{Abbott:2020gyp}.

We then combine all of the events using Eq.~\eqref{eq:product} to give the likelihood across all events. From this likelihood, we can then compute the posterior $p\left( \kappa \mid \left\{ D_{\mathrm{GW},j} \mid j = 1, \ldots, M_\mathrm{obs} \right\} \right)$ using a flat prior on $\kappa$, normalizing to integrate to 1.

We show the resulting posterior on $\kappa$ for GWTC-2 in Fig.~\ref{fig:p-kappa_distance}. We see that we do not get an informative constraint on $\kappa$ from O1-O2, as in the analysis presented in Sec.~\ref{sec:observed_constraint}. However, adding in O3a, we can get a constraint consistent with $\kappa = 0$.

Since $\kappa = 0$ corresponds to GR, by comparing the value of the posterior to the prior at $\kappa = 0$, we can obtain an \textit{evidence} for GR. While we see that for O1-O2 we effectively recover the prior value at $\kappa = 0$, giving us no information, in the case of the simulated detections, we can recover informative evidence for GR. However, for O3a and all of the detections, the result does give a constraint around $\kappa = 0$.

Note that this analysis requires that $\kappa d_C \ll 1$, and thus we must limit the values of $\kappa$ considered consistent with our events with large comoving distances for our analysis to be valid; going beyond linear order in the above relations is possible, but the solutions for $c_o\left( c, d_{L,o} \right)$ become multi-valued, significantly complicated the analysis. The events with confident constraints on inclination angle shown in Fig.~\ref{fig:CosIotaPosterior} are at redshifts of $0.05 \lesssim z \lesssim 0.38$. GWTC-2 does contain events at redshifts up to $z = 1$~\cite{Abbott:2020gyp}, but the inclination measurements from these events are uninformative. For future observations, however, we have to be cautious of the $\kappa d_C \ll 1$ requirement when bounding $\kappa$ with events at large redshifts in order for the linear analysis to remain valid.

We fit a Gaussian to the computed distribution on $\kappa$ (for all of the gravitational wave events) in Fig.~\ref{fig:p-kappa_distance}, finding a mean of $-0.035\textrm{ Gpc}^{-1}$, and a standard deviation of
$\sigma = 0.4 \textrm{ Gpc}^{-1}$. This value of $\sigma$ is larger than the width of the prior support we impose to satisfy the $\kappa d_C \ll 1$ constraint. We can estimate, however, how many future detections it will take for $\sigma$ to lie inside of the prior. For \textit{the same distance distribution of observed sources}, $\sigma$ will decrease by a factor of $\sqrt{N}$ for $N$ more detections. For $\sigma$ to decrease by a factor of two from $0.4 \textrm{ Gpc}^{-1}$ to $0.2 \textrm{ Gpc}^{-1}$, we thus require $N \sim 30$ more \textit{informative} events.

However, for future gravitational wave detections, we know that we will be able to observe further distances, which will affect the number of detections and hence the behavior of $\sigma$. Specifically, the rate at which we observe new events increases with distance $d_C$ as $d_C^3$ (since the overall observable volume increases). Thus, $\sigma$ will decrease with distance as $d_C^{-3/2}$.\footnote{Here we make the assumption that $\sigma$ is otherwise independent of distance, conservatively ignoring the fact that events that are further can give larger constraints on amplitude birefringence, and assuming that the inclination angle can be measured with similar accuracy at various distances.}  This increased distance, however, will decrease the allowed value of $\kappa$ (from the constraint $\kappa d_C \ll 1$) by a factor of $d_C^{-1}$. Thus, as the observable distance increases, $\sigma$, the variance on the measured $\kappa$, will decrease faster than the prior  on the allowed values of $\kappa$. Hence, in time, we will be able to make a more precise and valid measurement of $\kappa$.

\section{Uninformative inclination distributions}
\label{sec:uninformative_appendix}

In order the quantify the amount of information about $\kappa$ contained in the GWTC-2 detections, we must compare the results (whether qualitatively or quantitatively through a Kullback-Leibler divergence) to the distribution on $\kappa$ that we would get from detections that are completely uninformative about $\cos \iota$.  Of course, such uninformative measurements must generate a posterior for $\kappa$ that is equal to the prior (that is, they must generate a flat likelihood function); but it is an interesting test for any practical inference method that it satisfies this condition.

To generate such a test for our methods here, we produce an uninformative distribution on $\cos \iota$ for all detections.  We generate $N_\mathrm{samp}$ mock samples from a distribution that is $\mathcal{U} [-1, 1]$. We can then take the ensemble of $N_\mathrm{det}$ such detections and compute a likelihood distribution on $\Delta$ using the procedure in Sec.~\ref{sec:observational_effects}, following with a computation of $\kappa$.

However, when generating these samples, we must be careful about the fact that we are considering an \textit{uninformative} distribution. For each detection, we obtain a certain amount of Poisson noise given that we only have $N_\mathrm{samp}$ discrete samples. Naively, one would expect these Poisson fluctuations to cancel one another out as we accumulate more detections, converging to some `true value'. However, because each successive uninformative `measurement' of $\cos \iota \in \mathcal{U} [-1, 1]$ offers no new information, there is no such sense of convergence. Instead, the detections essentially result in a random walk in the slope of the likelihood with the $\Delta$ parameter.
We compute a log-likelihood distribution on $\Delta$ over all of the detections using
\begin{align}
\label{eq:deltaLL}
    \log \mathcal{L} (\Delta) = \sum_\mathrm{Detections} \log \left[N_- \frac{(1 - \Delta)}{2 N_\mathrm{samp}} + N_+\frac{(1 + \Delta)}{2 N_\mathrm{samp}} \right]
\end{align}
where for each detection, $N_-$ is the number of samples with $\cos \iota < 0$ and $N_+ = N_\mathrm{samp} - N_-$ is the number of samples with $\cos \iota > 0$.

For a uniform distribution, we would expect to have $N_- = N_+ = \frac{1}{2}$, so let us write, to linear order, $N_- / N_\mathrm{samp} = \frac{1}{2} + \epsilon$ and $N_+ / N_\mathrm{samp} = \frac{1}{2} - \epsilon$.  For any particular detection, assuming $N_\mathrm{samp} \gg 1$, $\epsilon$ is approximately normally distributed with mean zero and standard deviation $1/\sqrt{N_\mathrm{samp}}$.  Eq.~\eqref{eq:deltaLL} then results in
\begin{align}
    \log \mathcal{L} (\Delta)  = \sum_\mathrm{Detections} \log \left[\frac{1}{2} - \Delta \epsilon \right]\,,
\end{align}
which for each detection results in a line with slope linearly dependent on $\epsilon$.  Summing the independent, normally-distributed random variables $\epsilon$ gives
\begin{equation}
  \log \mathcal{L} (\Delta) = \mathrm{const} - \Delta \sum_\mathrm{Detections} \epsilon.
\end{equation}
The sum of normally-distributed $\epsilon$ results in a random-walk for the slope of the likelihood with $\Delta$; the sum is, itself, normally-distributed with mean zero and standard deviation $\sqrt{N_\mathrm{det} / N_\mathrm{samp}}$.  In order to ensure that uninformative detections do not accumulate a significant slope in $\mathcal{L}(\Delta)$, we must ensure that
\begin{align}
\label{eq:UninformativeN}
    N_\mathrm{samp} \gg N_\mathrm{det}
\end{align}
and thus have a number of samples that is dependent on the number of detections in the uninformative case. Note that this is different from what we do in practice, where we assume that the gravitational wave events \textit{are} informative about $\cos \iota$ and hence $\Delta$, and we use a fixed number of samples (1024 in this study) from each posterior distribution in our calculations.

\begin{figure}
    \includegraphics[width=1.0\columnwidth]{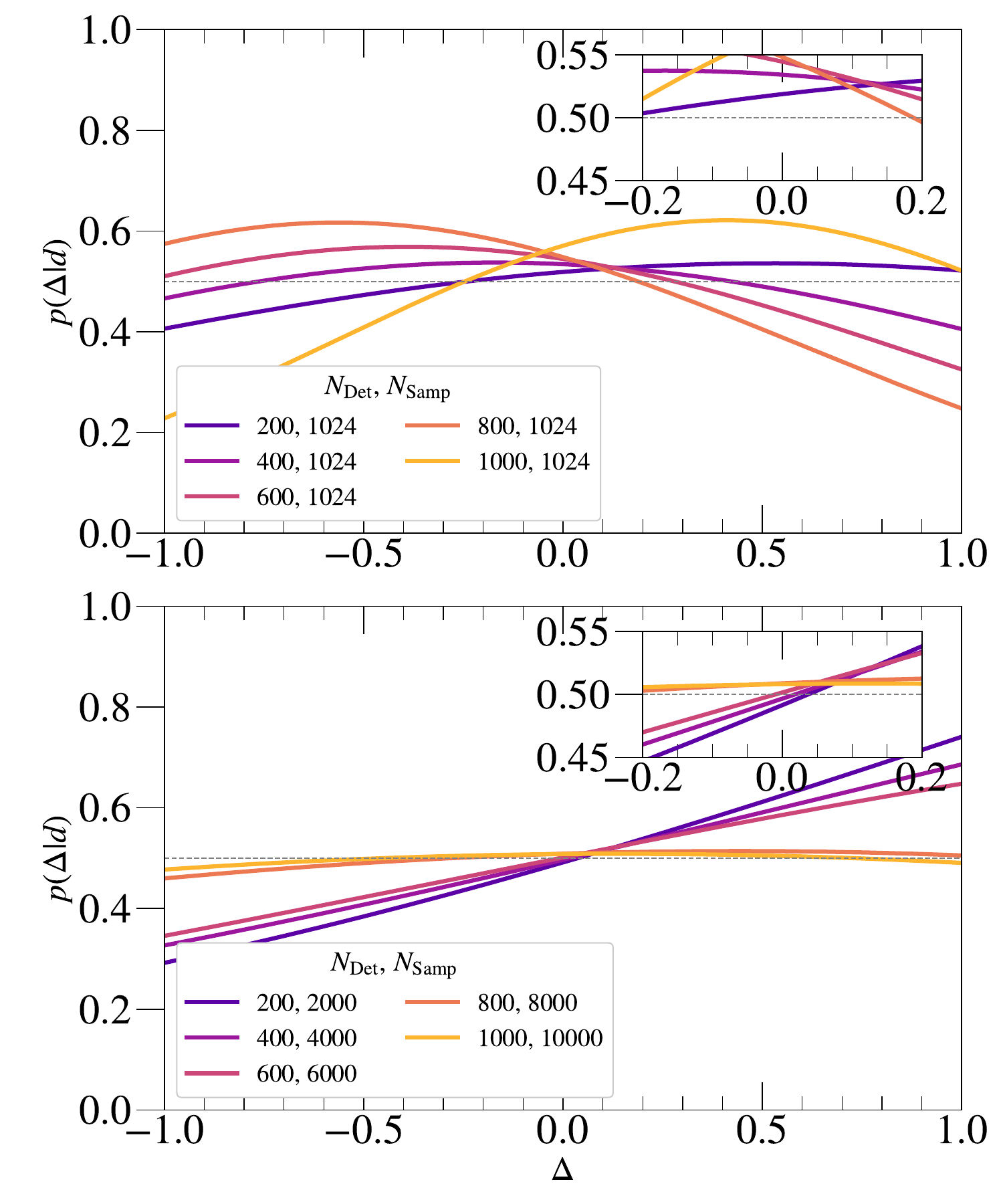}
  \caption{Posterior distribution on $\Delta$ computed from \textit{uninformative} distributions of $\cos \iota \in \mathcal{U} [-1, 1]$ for each detection. Each solid curve corresponds to the combined posterior distribution on $\Delta$ for the given number of detections, and the number of samples for each detection. The top panel corresponds to using a constant number of samples for each $N_\mathrm{det}$, which does \textit{not} converge to the expected flat distribution on $\Delta$ (dashed grey line) with increasing detections. The bottom panel, however, shows the case where $N_\mathrm{samp}$ changes with $N_\mathrm{det}$ to satisfy the criterion in Eq.~\eqref{eq:UninformativeN}, indeed showing convergence to the expected flat distribution.  The slope of the posterior is, in each case, comparable to $\sqrt{N_\mathrm{det} / N_\mathrm{samp}}$.
  }
  \label{fig:DeltaFluctuations}
\end{figure}

We can see the outcome of this in Fig.~\ref{fig:DeltaFluctuations}, where we plot the resulting distribution on $\Delta$ from uninformative samples with and without imposing the criterion in Eq.~\eqref{eq:UninformativeN}, where we obtain convergence to a flat distribution when we satisfy the criterion.

\section{Derivation of $\zeta(\eta)$ for dark-energy dominated universe}
\label{sec:zeta_appendix}

We now work through the derivation of $\zeta(\eta)$ (cf. Eq.~\eqref{eq:hZeta}) for a dark-energy dominated universe. We follow the steps of~\cite{Alexander:2007kv}, which computed $\zeta(\eta)$ for a matter-dominated universe.

We work in units of conformal time $\eta$, with $[\eta] = L^0$, and where $\eta = 1$ corresponds to present-day. The scale factor $a$ has units of $[a] = L$. The conformal time and proper time $t$ are related as $dt = a d \eta$. We use notation for derivatives $\dot{f} = \pd_t f$ and $f' = \pd_\eta f$. $H \equiv \dot{a}/a$ is the Hubble parameter, with $[H] = L^{-1}$, and $\mathcal{H} \equiv a'/a$ is the conformal Hubble parameter with dimensions $[\mathcal{H}] = L^0$. Quantities with subscript $0$, such as $\{a_0, H_0, \mathcal{H}_0\}$, refer to present-day values of the parameters. As stated before, the CS scalar field $\vartheta$ has dimensions of $[\vartheta] = L^2$, for the choice of $\alpha = \kappa$ for the CS coupling constant (cf. Eq.~\eqref{eq:Action}). We set $G = c = 1$ for this calculation.

Let us assume that right and left polarized gravitational waves have the following profile (cf. Eq. 189 in~\cite{Alexander:2009tp}),
\begin{align}
\label{eq:GWAnsatz}
    h_\RL = A (1 + \lambda_\RL \cos \iota)^2 \exp [-i (\phi_0 + \Delta \phi_\RL)]\,,
\end{align}
where $\iota$ is the inclination angle between the angular momentum of the source and the observer's line of sight, and $A$ is an amplitude dependent on parameters of the source that is the same for both polarizations. The quantity $\lambda_\mathrm{R} = +1$ for right-handed polarizations, and $\lambda_\mathrm{L} = -1$ for left-handed polarizations. The quantity $\phi_0$ is the gravitational wave phase as given by GR, and $\Delta \phi_\RL$ is the CS modification to the gravitational wave phase. Let us write the total phase as
\begin{align}
    \phi_\RL(\eta) = \phi_0(\eta) + \Delta \phi_\RL(\eta)\,,
\end{align}

With the profile in Eq.~\eqref{eq:GWAnsatz}, the ratio between the right and left polarized strain becomes
\begin{align}
\label{eq:hRhL}
    \frac{h_\mathrm{R}}{h_\mathrm{L}} = \frac{(1 + \cos \iota)^2}{(1 - \cos \iota)^2} \exp [-i (\Delta \phi_\mathrm{R} - \Delta \phi_\mathrm{L})]\,.
\end{align}
It is the quantity
\begin{align}
    \Delta \phi_\mathrm{R} - \Delta \phi_\mathrm{L}
\end{align}
    that we are thus interested in computing, and which is related to $\zeta$ (cf. Eq.~\eqref{eq:hZeta}) as
\begin{align}
\label{eq:ZetaDefinition}
    \frac{2 k}{H_0} \zeta = -i (\Delta \phi_\mathrm{R} - \Delta \phi_\mathrm{L})\,.
\end{align}

The standard linearized Einstein equations for metric perturbations in a Friedmann-Lema\^{i}tre-Robertson-Walker (FLRW) universe are modified through the inclusion of CS coupling to a scalar field. The equation for the phase of circularly polarized modes thus takes the form (cf. Sec. 2.B in~\cite{Alexander:2009tp} for a full derivation)
\begin{align}
\label{eq:AY184}
    &\left[ i\phi_\RL '' + (\phi_\RL ')^2 + \He' + \He^2 - \kappa^2 \right] \left(1 - \frac{\lambda_\RL \kappa \vartheta '}{a^2} \right) \\
    \nn & \quad = \frac{i \lambda_\RL \kappa}{a^2}(\vartheta'' - 2\He  \vartheta')(\phi_\RL' - i \He) \,,
\end{align}
where $\kappa$ is the co-moving wave-number with units $[\kappa] = L^0$. For ease of notation, let us drop the $\RL$ subscript and focus on a polarization with a generic $\lambda \in \{-1, 1\}$.

Following~\cite{Alexander:2007kv}, we put Eq.~\eqref{eq:AY184} in terms of a host of other variables, namely
\begin{align}
\label{eq:AYQuantities}
    &\;\;y \equiv \frac{\phi'}{k} \;\; \;\;\gamma \equiv \frac{\He_0}{\kappa}  \;\; \;\;\Gamma \equiv \frac{\He}{\He_0} \\
    \nn &\;\;\delta \equiv \frac{\He_0'}{\kappa^2}  \;\; \;\;\Delta \equiv \frac{\He'}{\He_0'} \;\; \;\; \epsilon = \frac{\vartheta_0''}{a_0^2} \\
    \nn &\;\; \zeta \equiv \frac{\kappa \vartheta_0'}{a_0^2} \;\; \;\; E \equiv \frac{\vartheta''}{a^2 \epsilon} \;\; \;\; Z \equiv \frac{\kappa \vartheta'}{a^2 \zeta}\,.
\end{align}
Eq.~\eqref{eq:AY184} thus becomes
\begin{align}
\label{eq:AYdY}
     \frac{y'}{\kappa} + i(1 - \gamma^2 \Gamma^2 - \delta \Delta - y^2) = \frac{\lambda (\epsilon E - 2 \gamma \zeta \Gamma Z)}{1 - \lambda \zeta Z} (y - i\gamma \Gamma)\,.
\end{align}

Thus far, nothing has been assumed about the scale factor or matter-energy content of the FLRW universe. Let us assume, however, following~\cite{Alexander:2007kv} that $\vartheta$ and $\He$ evolve on cosmological timescales (with $f' \sim \He f$), and so
\begin{align}
    \epsilon^2 \sim (\gamma \zeta)^2  \ll \gamma^2 \sim \delta\,.
\end{align}

Then, we can say that all of the terms with factors of $\epsilon$ and $\gamma \zeta$ are perturbations, and hence we can write the solution to Eq.~\eqref{eq:AYdY} as
\begin{align}
    y = y_0 + \epsilon y_{0,1} + \gamma \zeta y_{1,0} + \ldots\,,
\end{align}
where $y_0$ is the value of $y$ obtained from pure GR (setting $\vartheta = 0$ in Eq.~\eqref{eq:AYdY}), and $\{\epsilon, \gamma, \zeta\}$ are given in Eq.~\eqref{eq:AYQuantities}.

Next, we require that the perturbations \textit{vanish at some initial conformal time $\eta_i$}, we obtain that (cf. Eq. 2.23 in~\cite{Alexander:2009tp})
\begin{align}
    y_{0,1}(\eta) &= \lambda \mathcal{Y}[E](\eta) \,, \\
    y_{1,0}(\eta) &= -2 \lambda \mathcal{Y}[\Gamma Z] (\eta)\,,
\end{align}
where $\{E, \Gamma, Z\}$ are functions of $\vartheta$ given in Eq.~\eqref{eq:AYQuantities}, and
\begin{align}
    \mathcal{Y}[g](\eta) \equiv \kappa e^{2i\phi_0(\eta)} \int_{\eta_i}^\eta dx e^{-2i\phi_0(x)} y_0(x) g(x)\,,
\end{align}
for some function $g(\eta)$, where $\phi_0(\eta)$ is the gravitational wave phase from pure GR (obtained from solving Eq.~\eqref{eq:AY184} with $\vartheta = 0$).

The CS correction to the accumulated phase as the wave propagates from $\eta_i$ to $\eta$ (cf. Eq. 2.24 in~\cite{Alexander:2009tp}) is thus
\begin{align}
\label{eq:DeltaPhiIntegral}
    \Delta \phi(\eta_i, \eta) = \kappa \lambda \int_{\eta_i}^{\eta} d \eta \{ \epsilon \mathcal{Y}[E](\eta) - 2 \gamma \zeta \mathcal{Y}[\Gamma Z](\eta) \}\,.
\end{align}

To summarize, our goal is to integrate Eq.~\eqref{eq:DeltaPhiIntegral} to obtain the CS modification to the gravitational wave phase, which will allow us to compute the ratio between right and left polarized stain modes for a given $\vartheta$, as expressed in Eq.~\eqref{eq:hRhL}.

If we assume that $\gamma \ll 1$ (which is justified for the LIGO frequency range) then the function $\mathcal{Y}[g]$ (for some function $g[\eta]$) has the asymptotic expansion (cf. Eq. 2.25 in~\cite{Alexander:2007kv})
\begin{align}
\label{eq:YAsymptotic}
    \mathcal{Y}[g](\eta) \sim \frac{i e^{2 i \phi_0(\eta)}}{2} \left[ e^{-2i\phi_0 (\eta)} \sum_{\ell = 0}^n \left(\frac{1}{2ik}\right)^\ell + \left(\frac{1}{y_0} \frac{d}{d\eta} \right)^\ell g \right]^\eta_{\eta_i}\,.
\end{align}
We will follow~\cite{Alexander:2007kv} in going to order $\ell = 0$ in this calculation, giving
\begin{align}
    \nn \mathcal{Y}[g](\eta) &\sim \frac{i e^{2i\phi_0(\eta)}}{2} \left(e^{-2i\phi_0(\eta)}g(\eta) - e^{-2i\phi_0(\eta_i)}g(\eta_i) \right) \\
    \label{eq:YAsymptoticZero}
    &= \frac{i}{2} g(\eta) - \frac{i}{2} e^{2i(\phi_0(\eta) - \phi_0(\eta_i))}g(\eta_i)\,.
\end{align}

Now our calculation diverges from that in~\cite{Alexander:2009tp}, as we work in a dark-energy dominated (rather than matter-dominated) universe, with scale factor
\begin{align}
    a(t) = a_0 e^{H_0 t}\,.
\end{align}
Working in units of conformal time, we obtain
\begin{align}
    \eta(t) &= -\frac{1}{a_0 H_0} e^{-H_0 t} \,,\\
    t(\eta) &= \frac{\log\left(-\frac{1}{a_0 H_0 \eta}\right)}{H_0} \,,
\end{align}
which gives
\begin{align}
a(\eta) &= -\frac{1}{H_0 \eta}\,.
\end{align}

With the convention of $\eta = 1$ corresponding to present day, we obtain
\begin{align}
    a_0 &= -\frac{1}{H_0}\,, \;\;\;\; a'(\eta) = \frac{1}{H_0 \eta^2} \,, \\
    \He &\equiv \frac{a'(\eta)}{a(\eta)} = \frac{1}{\eta}\,, \;\;\;\; \He_0 = 1 \,, \\
    \He' &= -\frac{1}{\eta^2}\,, \;\;\;\;
    \He'_0 = -1 \,.
\end{align}
Now, we can compute all of the quantities in Eq.~\eqref{eq:AYQuantities} for a dark-energy dominated universe as
\begin{align}
\label{eq:AYQuantitiesDarkEnergy}
    &\;\;y \equiv \frac{\phi'}{\kappa} \;\;\;\;\gamma \equiv \frac{1}{\kappa}  \;\;\;\;\Gamma \equiv \frac{1}{\eta} \\
    \nn &\;\;\delta \equiv \frac{-1}{\kappa^2}  \;\;\;\;\Delta \equiv \frac{1}{\eta^2}\;\;\;\; \epsilon = H_0^2 \vartheta_0'' \\
    \nn &\;\; \zeta \equiv \kappa \vartheta_0' H_0^2\;\;\;\; E \equiv \eta^2 \frac{\vartheta'' }{\vartheta_0''} \;\;\;\; Z \equiv \eta^2 \frac{\vartheta' }{\vartheta_0'}\,.
\end{align}

Our aim is thus to evaluate Eq.~\eqref{eq:DeltaPhiIntegral} to obtain the CS correction to the phase, $\Delta \phi$.
Now, we must first obtain $y_0$, the value of $\phi_0'/k$ without a perturbation. Thus, we solve Eq.~\eqref{eq:AYdY} with zero RHS to give
\begin{align}
    \nn &\frac{y_0'}{k} + i(1 - \gamma^2 \Gamma^2 - \delta \Delta - y^2) = 0 \\
    \nn &\frac{y_0'}{k} + i(1 - \frac{1}{\kappa^2 \eta^2} - \frac{-1}{\kappa^2 \eta^2} - y^2) = 0\\
    &\frac{y_0'}{k} + i(1 - y^2) = 0 \,,
\end{align}
which gives solutions of the form
\begin{align}
    y_0 = -i \tan(\kappa \eta -i C_0)\,,
\end{align}
where $C_0$ is a constant of integration that we will leave unspecified for now.

Integrating
\begin{align}
    \phi_0' = \kappa y_0 \,,
\end{align}
we obtain
\begin{align}
\label{eq:Phi0Solution}
    \phi_0 (\eta) = C_1 + i \log(\cosh(C_0 - i \kappa \eta))
\end{align}
we can freely set $C_1 = 0$ since we are interested in the difference between two values of $\phi_0$.

Now, let us find $\Delta \phi$, the CS phase accumulated by the perturbations using Eq.~\eqref{eq:DeltaPhiIntegral}. Using the form of $\mathcal{Y}[g]$ from Eq.~\eqref{eq:YAsymptoticZero},  and the solution in Eq.~\eqref{eq:Phi0Solution}, we compute
\begin{align}
& \phi_0(\eta) - \phi_0(\eta_i) = \\
\nn & \quad -i \log(\cosh(C_0 - i\kappa\eta)) + i \log(\cosh(C_0 - i\kappa\eta_i))
\end{align}
which gives
\begin{align}
    &e^{2i(\phi_0(\eta) - \phi_0(\eta_i))} = \frac{\cosh(C_0 - i\kappa \eta)^2}{\cosh(C_0 - i\kappa\eta_0)^2}
\end{align}
Thus, we have
\begin{align}
    \nn \epsilon \mathcal{Y}[E](\eta) &= \frac{i}{2} H_0^2 \vartheta_0'' \left(\eta^2 \frac{\vartheta''}{\vartheta_0''} - \eta_i^2 \frac{\vartheta_i''}{\vartheta_0''} \frac{\cosh(C_0 - i\kappa \eta)^2}{\cosh(C_0 - i\kappa\eta_0)^2} \right) \\
    &= \frac{i}{2} H_0^2  \left(\eta^2 \vartheta''- \eta_i^2 \vartheta_i'' \frac{\cosh(C_0 - i\kappa \eta)^2}{\cosh(C_0 - i\kappa\eta_0)^2}\right)
\end{align}
and similarly
\begin{align}
    \nn \gamma \zeta \mathcal{Y}[\Gamma Z](\eta) &= \frac{i}{2}\vartheta_0' H_0^2 \left(\eta \frac{\vartheta'}{\vartheta_0'} - \eta_i \frac{\vartheta_i'}{\vartheta_0'} \frac{\cosh(C_0 - i\kappa \eta)^2}{\cosh(C_0 - i\kappa\eta_0)^2} \right) \\
    &= \frac{i}{2}H_0^2 \left(\eta \vartheta' - \eta_i \vartheta_i'\frac{\cosh(C_0 - i\kappa \eta)^2}{\cosh(C_0 - i\kappa\eta_0)^2} \right) \,.
\end{align}
Let us follow the logic below Eq. 3.4 of~\cite{Alexander:2009tp} to drop the oscillatory pieces, thus obtaining the overall integral from Eq.~\eqref{eq:DeltaPhiIntegral} of
\begin{align}
    \Delta \phi_\RL \sim i \frac{\kappa}{2} \lambda_\RL H_0^2 \int_\eta^1 \left( \eta^2 \vartheta''(\eta)  - 2 \eta \vartheta'(\eta) \right) d \eta\,,
\end{align}
where we have reintroduced the $\RL$ notation.

Following Eq.~\eqref{eq:ZetaDefinition}, we have
\begin{align}
    \frac{2 k}{H_0} \zeta = -i (\Delta \phi_\RR - \Delta \phi_\LL)
\end{align}
and thus, using $\lambda_\RR - \lambda_\LL = 2$, we obtain
\begin{align}
    \frac{2 k}{H_0} \zeta = \kappa H_0^2 \int_\eta^1 \left( \eta^2 \vartheta''(\eta)  - 2 \eta \vartheta'(\eta) \right) d \eta\,.
\end{align}
Writing $\kappa = k_0 / H_0$ (cf. Eq. 3.8 in~\cite{Alexander:2007kv}), we obtain,
\begin{align}
\label{eq:ZetaFinal}
    \zeta = \frac{H_0^2}{2} \int_\eta^1 \left( \eta^2 \vartheta''(\eta)  - 2 \eta \vartheta'(\eta) \right) d \eta\,.
\end{align}
Eq.~\eqref{eq:ZetaFinal} precisely gives us $\zeta$ for a dark-energy dominated universe. Let us double-check the units. In Eq.~\eqref{eq:ZetaDefinition}, $\zeta$ must be dimensionless. In this study, $[\vartheta] = L^2$ and $[H_0] = L^{-1}$, so indeed $[\zeta] = L^0$.

\bibliography{biblio}
\end{document}